\documentclass[a4paper, 12pt]{article}
\usepackage{amsfonts, amsthm, amsmath}
\usepackage[utf8]{inputenc}

\usepackage{amssymb}
\usepackage{graphicx}
\usepackage{xcolor}
\usepackage{cite}
\usepackage{qtree}
\usepackage{float}
\usepackage{lscape}
\numberwithin{equation}{section}

\newtheorem{theorem}{Theorem}

\allowdisplaybreaks

\title{ {\bf
Superintegrability in the interaction of two particles with spin}}

\author{\\
        {\bf O. O\u{g}ulcan Tuncer}\thanks{E-mail address:
        otuncer@hacettepe.edu.tr}\,, \,
        { \bf \.{I}smet Yurdu\c{s}en}\thanks{E-mail address:
       yurdusen@hacettepe.edu.tr}
 \\
  \\Department of Mathematics, Hacettepe University,
                    \\ 06800 Beytepe, Ankara, Turkey}

\date{\today}
\newpage
\begin{document}
\setlength{\baselineskip}{24pt} 
\maketitle
\setlength{\baselineskip}{7mm}
\begin{abstract}
We initiate a research program for the systematic investigation of quantum superintegrable systems involving the interaction of two non-relativistic particles with spin $1/2$ moving in the three-dimensional Euclidean space. In this paper, we focus specifically on such superintegrable systems that allow additional scalar integrals of motion, linear in the momenta. We first identify specific potentials appearing in the Hamiltonian that should be excluded from the analysis, as they can be immediately derived through gauge transformations of the natural Hamiltonian. Next, we construct the most general symmetric scalar operator and derive the determining equations from its commutation with the Hamiltonian. Solving these equations, we obtain 30 new superintegrable systems with spin, along with their corresponding first-order scalar integrals of motion.
\end{abstract}
Keywords: superintegrable systems, spin, scalar integrals of motion, nucleon-nucleon interaction, gauge transformation \medskip \\
PACS numbers: 02.30.Ik, 03.65.-w, 11.30.-j, 13.75.Cs

\section{Introduction}
The aim of this article is to initiate a research program for performing a systematic study of  superintegrable systems in the interaction of two particles with spin. This system can be physically interpreted e.g. as a nucleon-nucleon interaction. A similar research program was previously conducted by constraining the spin of one of the particles to zero. Before delving into the details of the problem of this paper, we shall briefly review the basics and related works. 

In classical mechanics, a Hamiltonian system with $n$ degrees of freedom is called \textit{integrable}, if the system allows $n$ functionally independent integrals of motion in involution including the Hamiltonian. Such an integrable system is called \textit{superintegrable} if the system admits at least one and at most $n-1$ additional integrals of motion. A superintegrable system with one additional integral of motion is called \textit{minimally superintegrable}, while a superintegrable system with $n-1$ additional integrals of motion is called \textit{maximally superintegrable}. Note that the total set of integrals (at most $2n-1$) must be functionally independent, but the additional integrals are not necessarily in involution, either among themselves or with the already existing $n$ integrals (except the Hamiltonian). The quantum counterparts of these definitions are quite similar, however the integrals of motion are now well-defined linear quantum mechanical operators and the functionally independence is replaced by the algebraically independence.

For natural Hamiltonians of the form
\begin{eqnarray}
	H=-\frac{\hbar^2}{2} \Delta + V_0(\vec{x})\,,
	\label{spinlesshamiltonian}
\end{eqnarray}
the systematic investigation of classical and quantum superintegrable systems with integrals that are first- and second-order polynomials in the momenta dates back to the 1960s \cite{Fris, Winternitz.b, Makarov}. First-order integrals are associated with geometrical symmetries of the potential, while 
second-order integrals are directly related to the separation 
of variables in the Schr\"{o}dinger equation 
or Hamilton-Jacobi equation in the classical case 
\cite{Makarov, Fris, Winternitz.b, Evans.a, Evans.b, Miller.a}. For the case of spherically symmetric natural Hamiltonian, there are two famous superintegrable systems: the Kepler or Coulomb system (where $V_0(r)=\frac{\alpha}{r}$) and the Harmonic oscillator (where $V_0=\omega r^2$). These systems are characterized by the fact that all finite classical trajectories in the system are periodic. Indeed, from Bertrand’s theorem \cite{Bertrand} we already know that these are the only two spherically symmetric potentials in which all bounded trajectories are closed. By relaxing the condition of spherical symmetry in the potential, one can explore a variety of new possibilities, such as the anisotropic harmonic oscillator with rational ratio of frequencies \cite{Jauch}.

Second-order superintegrability has been investigated in two and three dimensional spaces of constant and nonconstant curvature \cite{Kalnins.a,Grosche1,Grosche2,Grosche3,Kalnins.d,Kalnins.e,Kalnins.f} and also studied in $n$ dimensions \cite{Rodriguez,Kalnins.h,Kalnins.i}. For higher-order superintegrability see \cite{Gravel.a,Gravel.b,Tremblay.b,Tremblay.a,Marquette.b,Tremblay.c,Post,Quesne,Marquette.c,Popper,PostWinternitz:2015,MSW,AMJVPW2015,AMJVPWIY2018,AW,IanMproc2,EWY,Yurdusenconf,MPR,ELW}. The review article \cite{MillerPostWinternitz:2013} provides a comprehensive overview of all topics covered in the preceding discussion.

A generalization of the natural Hamiltonian \eqref{spinlesshamiltonian} is the Hamiltonian involving velocity dependent potentials. In particular, there are many studies on integrable and superintegrable Hamiltonian systems of a particle moving under the influence of magnetic field. See \cite{Dorizzi,Berube} and the recent papers \cite{Libor1,Libor2,Libor3,Libor4,Libor5,Libor6}.


The journey of investigating and classifying superintegrable systems involving two non-relativistic particles, one with spin $s=1/2$ (e.g. a nucleon) and the other with spin $s=0$ (e.g. a pion), was started by \cite{Winternitz.c} in 2006, where additional integrals of motion are assumed to be first-order and the motion is restricted to  the Euclidean plane $E_2$. In \cite{Yurdusen2015}, a similar problem was considered by adding the second-order integrals of motion to the already existing first-order integrals of motion. The same problem was also carried out for particles moving in the three-dimensional Euclidean space $E_3$ with additional integrals of motion, restricted to first-order ones \cite{WYproceeding, wy3} and second-order ones \cite{DWY, YTW}. In those cases the following form of the Hamiltonian was considered
\begin{eqnarray}
	H = -\frac{\hbar^2}{2} \Delta + V_0(\vec{x}) + \frac{1}{2}\Big\{V_1(\vec{x}), 
	(\vec{\sigma}, \vec{L})\Big\}\,,
	\label{generalhamiltonianwithhsec1}
\end{eqnarray}
where the $V_1(\vec{x})$ term represents the spin-orbital interaction, $\vec{L}$ is the angular momentum operator, $\vec{\sigma}$ is the Pauli vector, and the curly bracket denotes an anticommutator. Assuming the potentials appearing in \eqref{generalhamiltonianwithhsec1} to be spherically symmetric, all the first- and second-order scalar, pseudo-scalar, vector, axial-vector, tensor, and pseudo-tensor integrals of motion were found. Moreover, the classification problem of the superintegrable systems for these cases was completed. 

Other than this systematic approach, previously there have been two more types of studies in the quest for superintegrability with spin. One of them involves a particle with spin interacting with an external field, \emph{e.g.} an electromagnetic one \cite{Pronko.a, Pronko.b, Nikitin.a, Nikitin.b, Nikitin.d, Nikitin.e, Nikitin.f}. The other one pertains to a spin $\frac{1}{2}$ particle interacting with a dyon \cite{DHoker} or with a self-dual monopoles \cite{Feher}. 


Of course, the next natural generalization of the Hamiltonian given in \eqref{generalhamiltonianwithhsec1} is to consider two non-relativistic particles, both with spin $1/2$. This setup may be considered physically richer than the previous case, where the spin of one particle was set to zero, as it could encompass all nucleon-nucleon interactions.
Consequently, we anticipate identifying a significantly greater number of superintegrable systems compared to the previous case. These systems may offer much more concrete physical interpretations, which forms the core motivation for this study.

This paper is organized in the following fashion. In Section 2 we begin with a brief review of how to derive the most general Hamiltonian for a system of two particles with spin, and we also provide the formulation of our problem. We have already mentioned that we shall completely classify all first-order superintegrable systems involving two non-relativistic particles with spin. However, in the analysis we can exclude some certain particular potentials from our discussion, because these potentials can be directly induced by applying a gauge transformation to the scalar Hamiltonian in \eqref{spinlesshamiltonian}. In Section 3 we derive these potentials that should be excluded from the classification analysis. In Section 4 after obtaining the most general symmetric first-order scalar operator, we derive the determining equations coming from the commutation of this operator with the most general Hamiltonian. Then we classify all superintegrable systems in the interaction of two particles with spin $1/2$, allowing additional first-order scalar integrals of motion. We close in Section 5 with
a summary of our results along with some future directions.

\section{Formulation of the problem}
We will deal with two interacting non-relativistic particles with coordinates $x_i$, momenta $p_i$, and Pauli spin operators $\sigma_i$ ($i=1,2$) and both of these particles are assumed to have spin $\frac{1}{2}$. In order to consider this problem, we need to know the Hamiltonian for the system. Hence we shall briefly discuss what restrictions should be imposed to derive the Hamiltonian for such a system. The problem of deriving a general form of non-relativistic potential between two nucleons was treated in 1941 by Eisenbud and Wigner \cite{EW}. They considered a potential that is limited by invariance conditions (invariance under rotations in space, invariance under reflections in space, invariance under time-reversal, and invariance under rotations in isobaric space) and that depends at most linearly on the relative momentum $p$. Hence it turned out that the general form of the nucleon-nucleon potential consists of central, spin-spin, tensor, and spin-orbit terms. In 1958, Okubo and Marshak \cite{OM} generalized the arguments of \cite{EW} and provided the most general velocity-dependent potential for the same problem by dropping the restriction that the relative momentum $p$ appears at most linearly and instead allowing all powers of $p$. 

We now give more details on the most general potential introduced by \cite{OM} by writing explicitly the restrictions on this potential. We begin by assuming charge independence of the two nucleon interaction. Note that we do not consider the isotopic-spin dependence throughout the paper, since the potential can simply be expressed as a linear combination of $1$ and $\tau_1\cdot\tau_2$ when the isotopic spin is concerned, where $\tau_j$ are the isotopic spin operators. Therefore, the potential can only depend on the coordinates $x^{(j)}$,  momenta $p^{(j)}$, and Pauli spin operators $\sigma_j$ ($j=1,2$). The following conditions are imposed on the potential: 
\begin{enumerate}
    \item[(i)] \textit{Translational invariance}: This condition means that the coordinates $x^{(j)}$ may appear in the potential only in the form $x=x^{(1)}-x^{(2)}$ which is called the relative separation.
    \item[(ii)] \textit{Galilean invariance}: This condition means that the potential depends on the momenta $p^{(j)}$ only through the relative momentum $p=\frac{1}{2}(p^{(1)}-p^{(2)})$. In other words, the potential is independent of the total momentum $p=p^{(1)}+p^{(2)}$.
\end{enumerate}
These two conditions will simply allow us to write $V\equiv V(\sigma_1, \sigma_2, x, p)$.
\begin{enumerate}
    \item[(iii)]  \textit{Permutation symmetry}: This condition simply means that the potential is not affected by interchanging two particles, that is, $V(\sigma_1, \sigma_2, x, p)=V(\sigma_2, \sigma_1, -x, -p)$.
    \item[(iv)] \textit{Rotational invariance}: This condition means that invariant functions in the potential may be functions of $x^2$, $p^2$, and $L^2=(x\times p)^2$.
    \item[(v)] \textit{Space reflection invariance}: This condition is equivalent to $V(\sigma_1, \sigma_2, x, p)=V(\sigma_1, \sigma_2, -x, -p)$. Applying (iii) to this condition, we have $V(\sigma_1, \sigma_2, x, p)=V(\sigma_2, \sigma_1, x, p)$.
    \item[(vi)] \textit{Time reversal invariance}: This condition implies that $V(\sigma_1, \sigma_2, x, p)=V^T(-\sigma_1, -\sigma_2, x, -p)$, where $V^T$ simply reverses the order of operators.
    \item[(v)] \textit{Hermiticity}: $V=V^\dagger$.
\end{enumerate}
Okubo and Marshak derived the most general potential relying on the conditions listed above. As for $\sigma$-dependence, the potential is separated into the zeroth, the first, and the second-order terms with respect to  $\sigma$. Hence the zeroth-order term that is the term without spin operators, say $V_c$, is a function of $x^2$, $p^2$, and $L^2=(x\times p)^2$. The first-order term with respect to the spin operators will be in the form $(S,L) V_{so}$, where $S=(\sigma_1+\sigma_2)/2$. The other possible forms like $(S, x)$ or $(S, p)$ are excluded due to the space reflection invariance of the potential. Finally the second-order terms with respect to the spin operators may involve the following four independent factors 
\begin{equation*}
    (\sigma_1,\sigma_2),\quad (\sigma_1,x)(\sigma_2,x),\quad (\sigma_1,p)(\sigma_2,p),\:\: \textrm{and}\:\: (\sigma_1,L)(\sigma_2,L).
\end{equation*}
Hence the most general potential is given by 
\begin{align}
    V=&V_c+V_{so}(\vec{S},\vec{L})+V_{ss}(\vec{\sigma}_1,\vec{\sigma}_2)+V_t S_{12}+V_{sm}(\vec{\sigma}_1,\vec{p})(\vec{\sigma}_2,\vec{p}) \nonumber\\
    &+V_{qso}\dfrac{1}{2}\big((\vec{\sigma}_1,\vec{L})(\vec{\sigma}_2,\vec{L})+(\vec{\sigma}_2,\vec{L})(\vec{\sigma}_1,\vec{L})\big), \label{eqgenpotential}
\end{align}
where $S_{12}=\dfrac{3}{r^2}(\vec{\sigma}_1,\vec{x})(\vec{\sigma}_2,\vec{x})-(\vec{\sigma}_1,\vec{\sigma}_2)$ is the tensor operator.
Therefore this potential consists of six terms: a central term $V_c$, a spin-orbit term $V_{so}$, a spin-spin term $V_{ss}$, a tensor term $V_t$, a spin-momentum term $V_{sm}$, and a quadratic spin-orbit term $V_{qso}$. Each individual potential $V_i$ appearing in the general potential $V$ is a function of $x^2, p^2$, and $L^2$.
Note that if we are interested only in the scattering matrix for two nucleons, where the energy must be conserved, then the spin-momentum term can be written as a linear combination of other terms and so we have just five independent terms in the potential $V$ \cite{Puzikov}. Hence, on the energy shell, only five of the six terms are independent and so the spin-momentum term can be eliminated. Besides, on the energy shell, it is sufficient that all $V_i$ appearing in the potential $V$ are assumed to be functions of $x^2$ and $L^2$, but not $p^2$. However, off the energy shell, the spin-momentum term should be considered as an independent term and the general potential given by \eqref{eqgenpotential} might be relevant.  

Set $r=\|\vec{x}\|$, which is the scalar distance. From now on we shall assume that the potentials appearing in the Hamiltonian are functions of only $r$, i.e., they are spherically symmetric. In this case the Hamiltonian we take care of is in the form
\begin{equation}\label{twospinHamiltonian}
	H=-\dfrac{\hbar^2}{2}\Delta+V,
\end{equation}
where
\begin{align*}
	V=&V_0(r)+V_1(r)\dfrac{1}{2}(\vec{\sigma}_1+\vec{\sigma}_2,\vec{L})+V_2(r)(\vec{\sigma}_1,\vec{\sigma}_2)+V_3(r)(\vec{x},\vec{\sigma}_1)(\vec{x},\vec{\sigma}_2)+V_4(r)(\vec{\sigma}_1,\vec{p})(\vec{\sigma}_2,\vec{p})\\
	&+V_5(r)\dfrac{1}{2}\big((\vec{\sigma}_1,\vec{L})(\vec{\sigma}_2,\vec{L})+(\vec{\sigma}_2,\vec{L})(\vec{\sigma}_1,\vec{L})\big)
\end{align*}
is given in the three-dimensional Euclidean space $E_3$. Note that $\vec{p}=(p_1,p_2,p_3)$ is the linear momentum, $\vec{L}=(L_1,L_2,L_3)$ is the angular momentum, and $\vec{\sigma}=(\sigma_x, \sigma_y, \sigma_z)$ is the Pauli vector. More specifically, we adopt the following standard notation
\begin{align*}
	&p_1=-\mathrm{i}\hbar\partial_x,\: p_2=-\mathrm{i}\hbar\partial_y,\: p_3=-\mathrm{i}\hbar\partial_z, \medskip\\
	& L_1=\mathrm{i}\hbar(z\partial_y-y\partial_z),\: L_2=\mathrm{i}\hbar(x\partial_z-z\partial_x),\: L_3=\mathrm{i}\hbar(y\partial_x-x\partial_y),\medskip \\
	&\sigma_x = 
	\left( \begin{array}{cc}
		0 & 1 \\
		1 & 0 \end{array} \right),\:
	\sigma_y = 
	\left( \begin{array}{cc}
		0 & -i \\
		i & 0 \end{array} \right),\: 
	\sigma_z = 
	\left( \begin{array}{cc}
		1 & 0 \\
		0 & -1 \end{array} \right),\medskip\\
	&\vec{\sigma}_1=\vec{\sigma}\otimes{I}_2;\quad\vec{\sigma}_2={I}_2\otimes\vec{\sigma},
\end{align*}
where $I_2$ is the $2\times 2$ identity matrix. Therefore, clearly each component of $\vec{\sigma}_1$ and $\vec{\sigma}_2$ is a $4\times 4$ matrix, so we can decompose the Hamiltonian given by \eqref{twospinHamiltonian} in terms of $4\times 4$ identity matrix $I_4$ and Dirac matrices $\sigma_{1,k}$, $\sigma_{2,k}$ for $k\in\{x,y,z\}$. However, for simplicity we drop the identity matrix from the Hamiltonian, whenever it is convenient. These Dirac matrices are explicitly given by
\[ \sigma_{1x} = 
\left( \begin{array}{cccc}
	0 & 0 & 1 & 0 \\
0& 0&	0 & 1 \\
1 & 0 & 0 & 0 \\
0 & 1 & 0 & 0
 \end{array} \right), \quad \sigma_{1y} = 
\left( \begin{array}{cccc}
0 & 0 & -i & 0 \\
0& 0&	0 & -i \\
i & 0 & 0 & 0 \\
0 & i & 0 & 0
\end{array} \right), \quad \sigma_{1z} = 
\left( \begin{array}{cccc}
1 & 0 & 0 & 0 \\
0& 1 &	0 & 0 \\
0 & 0 & -1 & 0 \\
0 & 0 & 0 & -1
\end{array} \right), \]
\[ \sigma_{2x} = 
\left( \begin{array}{cccc}
	0 & 1 & 0 & 0 \\
	1 & 0&	0 & 0 \\
	0 & 0 & 0 & 1 \\
	0 & 0 & 1 & 0
\end{array} \right), \quad \sigma_{2y} = 
\left( \begin{array}{cccc}
	0 & -i & 0 & 0 \\
	i & 0&	0 & 0 \\
	0 & 0 & 0 & -i \\
	0 & 0 & i & 0
\end{array} \right), \quad \sigma_{2z} = 
\left( \begin{array}{cccc}
	1 & 0 & 0 & 0 \\
	0& -1 &	0 & 0 \\
	0 & 0 & 1 & 0 \\
	0 & 0 & 0 & -1
\end{array} \right). \]
Our purpose in this paper is to find all the superintegrable potentials $V_i$ ($i=0,1,\ldots,5$) that admit additional scalar integrals of motion that are first-order matrix polynomials in the momenta. 
\section{Gauge transformation}
In this section, we decide what choices of potentials appearing in the Hamiltonian should be excluded from the analysis on a reason being derivable from a gauge transformation of a scalar Hamiltonian \eqref{spinlesshamiltonian}. A similar procedure was also applied to obtain redundant spin-orbit potentials for systems involving two particles with spin $1/2$ and $0$ \cite{wy3}. Now we will generalize this procedure to our case.

Consider the Hamiltonian for spinless particles (i.e., the scalar Hamiltonian) in \eqref{spinlesshamiltonian} multiplied by the $4\times 4$ identity matrix. The gauge transformation will be in the form
\begin{equation}\label{gauge1}
	\widetilde{H}=U^{-1}HU,
\end{equation}
and the corresponding transformation matrix is given by
\begin{equation}\label{gauge2}
	U=u\otimes u,\quad u\in \mathrm{U}(2).
\end{equation}
Notice that $U$ is also unitary matrix and so it is an element of $\mathrm{U}(4)$.
The general form of the matrix $u$ is 
\begin{equation}\label{gauge3}
	u=e^{\mathrm{i}\beta_4}\begin{pmatrix}
		e^{\mathrm{i}\beta_1}\cos(\beta_3) & e^{\mathrm{i}\beta_2}\sin(\beta_3)\medskip\\
		-e^{-\mathrm{i}\beta_2}\sin(\beta_3) & e^{-\mathrm{i}\beta_1}\cos(\beta_3)
	\end{pmatrix},
\end{equation}
where $\beta_j\,(j=1,2,3,4)$ are some real functions of $(x,y,z)$. Obviously, after applying the gauge transformation to the scalar Hamiltonian $H$, the resulting Hamiltonian $\widetilde{H}$ will be in the following form:
\begin{align}                   \tilde{H}=&H+\Gamma_0+\Gamma_1(\vec{\sigma}_1+\vec{\sigma}_2,\vec{L})+\Gamma_2(\vec{\sigma}_1,\vec{\sigma}_2)+\Gamma_3(\vec{x},\vec{\sigma}_1)(\vec{x},\vec{\sigma}_2)+\Gamma_4(\vec{\sigma}_1,\vec{p})(\vec{\sigma}_2,\vec{p})\nonumber\\
	&+\Gamma_5\big((\vec{\sigma}_1,\vec{L})(\vec{\sigma}_2,\vec{L})+(\vec{\sigma}_2,\vec{L})(\vec{\sigma}_1,\vec{L})\big), \label{gauge4}
\end{align}
where $\Gamma_j\,(j=0,1,\ldots,5)$ are real scalar functions of $(x,y,z)$.
Introducing the transformation matrix obtained by \eqref{gauge2} and \eqref{gauge3} into \eqref{gauge1} clearly gives a $4\times 4$ matrix whose entries incorporate at most second-order derivative operators. Then, an easy calculation shows that the only term having second-order derivative operators is $-\dfrac{\hbar^2}{2}\Delta$ which is already present in the scalar Hamiltonian $H$. This means that there is no second-order derivative term in $\widetilde{H}$ other than $-\dfrac{\hbar^2}{2}\Delta$. Therefore, $\Gamma_4$ and $\Gamma_5$ on the right-hand side of \eqref{gauge4} must vanish. Besides, it is clear that the right-hand side of \eqref{gauge4} has just one term including first-order derivatives which is $\Gamma_1(\vec{\sigma}_1+\vec{\sigma}_2,\vec{L})$. Hence, by equating the first-order derivative operators of \eqref{gauge1} and \eqref{gauge4}, we directly obtain 12 first-order PDEs for $\beta_j$ and $\Gamma_1$ given by 
\begin{align}
    & \beta_{4,x}=\beta_{4,y}=\beta_{4,z}=0, \label{gaugeeq1}\\
    & \hbar(\cos^2(\beta_3)\beta_{1,x}-\sin^2(\beta_3)\beta_{2,x})=-y\Gamma_1, \label{gaugeeq2}\\
    &  \hbar(\cos^2(\beta_3)\beta_{1,y}-\sin^2(\beta_3)\beta_{2,y})=x \Gamma_1,   \label{gaugeeq3}\\
    & \hbar(\cos^2(\beta_3)\beta_{1,z}-\sin^2(\beta_3)\beta_{2,z})=0, \label{gaugeeq4}\\
    & \hbar\left( \cos(\beta_1-\beta_2)\beta_{3,x}+\dfrac{1}{2}\sin(\beta_1-\beta_2)\sin(2\beta_3) ( \beta_{1,x}+\beta_{2,x}) \right)=z \Gamma_1, \label{gaugeeq5}\\
     & \hbar\left( -\sin(\beta_1-\beta_2)\beta_{3,y}+\dfrac{1}{2}\cos(\beta_1-\beta_2)\sin(2\beta_3) ( \beta_{1,y}+\beta_{2,y}) \right)=-z \Gamma_1, \label{gaugeeq6}\\
      & \hbar\left( \cos(\beta_1-\beta_2)\beta_{3,z}+\dfrac{1}{2}\sin(\beta_1-\beta_2)\sin(2\beta_3) ( \beta_{1,z}+\beta_{2,z}) \right)=-x \Gamma_1, \label{gaugeeq7}\\
      & \hbar\left( -\sin(\beta_1-\beta_2)\beta_{3,z}+\dfrac{1}{2}\cos(\beta_1-\beta_2)\sin(2\beta_3) ( \beta_{1,z}+\beta_{2,z}) \right)=y \Gamma_1, \label{gaugeeq8}\\
    & \sin(\beta_1-\beta_2) \beta_{3,x}-\dfrac{1}{2}\cos(\beta_1-\beta_2)\sin(2\beta_3) (\beta_{1,x}+\beta_{2,x})=0, \label{gaugeeq9}\\
    & \cos(\beta_1-\beta_2) \beta_{3,y}+\dfrac{1}{2}\sin(\beta_1-\beta_2)\sin(2\beta_3) (\beta_{1,y}+\beta_{2,y})=0.  \label{gaugeeq10}
\end{align}
These equations are similar to the ones given in \cite{wy3} and had already been solved to obtain gauge induced potentials for a system of two non-relativistic particles with spin $s=\frac{1}{2}$ and $s=0$. The solution of those 12 PDEs is
\begin{equation}
	\beta_1=\phi  \pm \pi + c\,, \quad \beta_2=c\,, \quad \beta_3=-\theta  \pm \frac{\pi}{2}\,, \quad \beta_4=0\,, \quad 0\le\theta\le \pi\,, \quad 0\le\phi<2\pi\,,  \nonumber 
	\label{specifiedgaugetrfmatrix}
\end{equation}
where $c$ is a constant, and $\theta$ and  $\phi$ are the spherical coordinates. We also have $\Gamma_1=\dfrac{\hbar}{r^2}$. Therefore, the gauge transformation matrix is found to be
\begin{equation}
	U=\begin{pmatrix}
		e^{2\mathrm{i}\phi}\sin^2\theta & -e^{\mathrm{i}\phi}\cos\theta\sin\theta & -e^{\mathrm{i}\phi}\cos\theta\sin\theta & \cos^2\theta \\
		e^{\mathrm{i}\phi}\cos\theta\sin\theta & \sin^2\theta & -\cos^2\theta & -e^{-\mathrm{i}\phi}\cos\theta\sin\theta\\
		e^{\mathrm{i}\phi}\cos\theta\sin\theta & -\cos^2\theta & \sin^2\theta & -e^{-\mathrm{i}\phi}\cos\theta\sin\theta \\
		\cos^2\theta & e^{-\mathrm{i}\phi}\cos\theta\sin\theta & e^{-\mathrm{i}\phi}\cos\theta\sin\theta & e^{-2\mathrm{i}\phi}\sin^2\theta
	\end{pmatrix}.
\end{equation}
Applying the gauge transformation to the scalar Hamiltonian $H$ based on this transformation matrix yields the Hamiltonian of the following form:
\begin{align}
	\tilde{H}=&U^{-1}\left(-\dfrac{\hbar^2}{2}\Delta+V_0(\vec{x}) \right)U \nonumber \\
	=&-\dfrac{\hbar^2}{2}\Delta+V_0(r)+\dfrac{2\hbar^2}{r^2}+\dfrac{\hbar}{r^2}(\vec{\sigma}_1+\vec{\sigma}_2,\vec{L})+\dfrac{\hbar^2}{r^2}(\vec{\sigma}_1,\vec{\sigma}_2)-\dfrac{\hbar^2}{r^4}(\vec{x},\vec{\sigma}_1)(\vec{x},\vec{\sigma}_2).
\end{align}
Hence the following set of potentials in the Hamiltonian \eqref{twospinHamiltonian} is gauge induced  
\begin{equation}\label{gaugepotentials}
  V_1=\dfrac{2\hbar}{r^2},\quad V_2=\dfrac{\hbar^2}{r^2},\quad V_3=-\dfrac{\hbar^2}{r^4},\quad V_4=0,\quad V_5=0.
\end{equation}
\subsection{Integrals for $V_0=V_0(r)$ and potentials in \eqref{gaugepotentials}}
We know that the set of potentials in \eqref{gaugepotentials} is gauge induced from the natural Hamiltonian \eqref{spinlesshamiltonian} each term of which is multiplied by the $4\times 4$ identity matrix. Therefore the integrals of motion for this case are found by applying gauge transformations to the integrals of this scalar Hamiltonian, i.e., $L_i$ and $S_i$. These integrals are
\begin{equation}
    \mathcal{J}_i=L_i+\hbar S_i,\quad \mathcal{S}_i=-\hbar S_i+\dfrac{2\hbar}{r^2}x_i (\vec{S},\vec{x}),
\end{equation}
and adhere to the following relations:
\begin{equation}
    [\mathcal{J}_i, \mathcal{J}_j]=\mathrm{i}\hbar \epsilon_{ijk}\mathcal{J}_k,\quad [\mathcal{S}_i, \mathcal{S}_j]=\mathrm{i}\hbar \epsilon_{ijk}\mathcal{S}_k, \quad [\mathcal{J}_i, \mathcal{S}_j]=\mathrm{i}\hbar \epsilon_{ijk}\mathcal{S}_k.
\end{equation}
Then the six-dimensional Lie algebra $\mathcal{L}$ is isomorphic to a direct sum of the algebra $o(3)$ with itself, that is,
\[ \mathcal{L} \sim o(3)\oplus o(3)=\{\vec{\mathcal{J}}-\vec{\mathcal{S}}\}\oplus \{ \vec{\mathcal{S}} \}. \]
\subsection{Integrals for $V_0=\dfrac{2\hbar^2}{r^2}$ and potentials in \eqref{gaugepotentials}}
These potentials are clearly gauge induced from the free Hamiltonian (the Hamiltonian \eqref{spinlesshamiltonian} with $V_0=0$). Then the integrals of motion for this case are obtained by applying gauge transformations to the integrals of motion of this free Hamiltonian, i.e., $L_i$, $S_i$, and $p_i$. The corresponding integrals are
\begin{equation}
      \mathcal{J}_i=L_i+\hbar S_i,\qquad \mathcal{P}_i=p_i-\dfrac{2\hbar}{r^2}\epsilon_{ikl}x_k S_l, \qquad \mathcal{S}_i=-\hbar S_i+\dfrac{2\hbar}{r^2}x_i (\vec{S},\vec{x}),  
\end{equation}
and they satisfy the following commutation relations:
\begin{align}
       & [\mathcal{J}_i-\mathcal{S}_i, \mathcal{S}_j]=0,\quad [\mathcal{P}_i, \mathcal{S}_j]=0, \quad [\mathcal{P}_i, \mathcal{P}_j]=0, \nonumber \\
       & [\mathcal{J}_i-\mathcal{S}_i, \mathcal{J}_j-\mathcal{S}_j]=\mathrm{i}\hbar \epsilon_{ijk}(\mathcal{J}_k-\mathcal{S}_k),\quad [\mathcal{J}_i-\mathcal{S}_i, \mathcal{P}_j]=\mathrm{i}\hbar \epsilon_{ijk}\mathcal{P}_k.
\end{align}
Thus we have a nine-dimensional Lie algebra, say $\mathcal{L}$, which is isomorphic to a direct sum of the Euclidean Lie algebra $e(3)$ with the algebra $o(3)$, that is,
\[ \mathcal{L} \sim e(3)\oplus o(3)=\{\vec{\mathcal{J}}-\vec{\mathcal{S}, \vec{\mathcal{P}}}\}\oplus \{ \vec{\mathcal{S}} \}. \]



\section{The scalar integrals of motion}
We shall first construct scalar operators in the direct product space of the vectors $(\vec{x}, \vec{p}, \vec{\sigma}_1, \vec{\sigma}_2)$.
Through the quantities $\vec{x}$, $\vec{p}$, $\vec{\sigma}_1$, and $\vec{\sigma}_2$, we can define ten independent directions given by
\begin{equation*}
	\{ \vec{x},\,\vec{p},\,\vec{L},\,\vec{\sigma}_1,\,\vec{\sigma}_2,\,\vec{\sigma}_1\times\vec{x},\,\vec{\sigma}_2\times\vec{x},\,\vec{\sigma}_1\times\vec{p},\,\vec{\sigma}_2\times\vec{p},\, {\vec{\sigma}_1\times\vec{\sigma}_2} \}.
\end{equation*}
Note that the first-order scalar integrals of motion can include arbitrary powers of $\vec{x}$, but are restricted to at most first powers of $\vec{p}$ and first powers of $\vec{\sigma}_1$ and $\vec{\sigma}_2$. Therefore all possible forms of the scalar operators are as follows: 
\begin{align*}
&	S_1=1,\quad S_2=(\vec{x},\vec{p}),\quad S_3=(\vec{\sigma}_1,\vec{\sigma}_2),\quad S_4=(\vec{x},\vec{\sigma}_1)(\vec{x},\vec{\sigma}_2),\medskip\\
&	S_5=(\vec{x},\vec{\sigma}_1)(\vec{\sigma}_2,\vec{p}),\quad S_6=(\vec{x},\vec{\sigma}_2)(\vec{\sigma}_1,\vec{p}),\quad S_7=(\vec{x},\vec{p})(\vec{\sigma}_1,\vec{\sigma}_2),\medskip \\
& S_8=(\vec{\sigma}_1,\vec{L}),\quad S_9=(\vec{\sigma}_2,\vec{L}),\quad S_{10}=(\vec{\sigma}_1,\vec{x})(\vec{\sigma}_2,\vec{x}) (\vec x,\vec p).
\end{align*}
Now we need to determine the most general form of a scalar operator. To do so, we will first multiply each term in the set of scalars by a weight function being a real function of $r$ and sum the results. This clearly gives us
\begin{equation*}
	\widetilde{Y}_s=\sum\limits_{j=1}^{10}f_j(r)S_j. 
\end{equation*}
Next we need to find the full symmetric form of this sum for the analysis of the commutation relation. This symmetrization process can be carried out by symmetrizing each term $f_j S_j$ ($j=1,\ldots,10)$ in the sum separately and then adding the results. To symmetrize these individual terms, we make use of a Mathematica code recently released by the authors of this paper \cite{YT}. This code simply outputs the full symmetric forms of a list of given operators in index form by calculating the permutations of each operator. We refer the interested reader to \cite{YT} for further details on this matter. After applying this symmetrization process, we obtain the full symmetric form of the most general scalar operator:
\begin{align*}
    Y_s=&\left( f_1(r)-\dfrac{\mathrm{i}\hbar}{2} \left( rf_2'(r)+3f_2(r) \right) \right)+\left( f_2(r)+f_7(r)\,(\vec\sigma_1, \vec\sigma_2)+f_{10}(r)(\vec{\sigma}_1,\vec{x})(\vec{\sigma}_2,\vec{x}) \right) (\vec x,\vec p) \\
    &+\left( f_3(r)-\dfrac{\mathrm{i}\hbar}{2}\left( f_5(r)+f_6(r)+rf_7'(r)+3 f_7(r) \right)  \right)(\vec{\sigma}_1,\vec{\sigma}_2)\\
    &+\left( f_4(r)-\dfrac{\mathrm{i}\hbar}{2}(f_5'(r)+f_6'(r)+rf'_{10}(r)+5f_{10}(r)) \right) (\vec{\sigma}_1,\vec{x})(\vec{\sigma}_2,\vec{x})+f_5(r)(\vec{\sigma}_1,\vec{x})(\vec{\sigma}_2,\vec{p})\\
    &+f_6(r)(\vec{\sigma}_2,\vec{x})(\vec{\sigma}_1,\vec{p})+f_8(r)(\vec{\sigma}_1,\vec{L})+f_9(r)(\vec{\sigma}_2,\vec{L}).
\end{align*}
Our aim now is to determine for which choices of $f_j$, $j=1,\ldots,10$ and $V_i$, $i=0,\ldots,5$ the most general symmetric scalar operator will commute with the Hamiltonian. Hence we shall consider the commutativity condition $[H,Y_s]=0$ to obtain the determining equations, and then we will solve these equations to find the appropriate potentials $V_i$ and functions $f_j$ so that $Y_s$ is an integral of motion.

Recall that $H$ and $Y_s$ are $4\times 4$ matrices. This means that the commutation $[H,Y_s]$ will also be a $4\times 4$ matrix. Therefore the commutativity condition is actually equivalent to all of the entries of this matrix being zero. Since there are 16 entries in this matrix, we will have a great number of determining equations even though some of these equations will coincide or can be derived from the others. Hence we will not present the complete list of these determining equations. Instead we will directly dive into the analysis by solving the determining equations coming from the highest-order derivatives.

Notice that the highest order derivatives in the commutation $[H,Y_s]$ are third-order ones. Equating the coefficients of these third-order derivatives to zero, we obtain the following set of determining equations:
\begin{align}
&(f_5-\mathrm{i}f_9)V_4=0,\quad (f_5-f_6)V_4=0, \quad (f_8-f_9)V_4=0, \quad (\mathrm{i}f_6+f_8)V_4=0, \nonumber \\
&(f_5+f_6)V_4=0,\quad (f_8+f_9)V_4=0,\quad f_{10}V_4=0. \label{deteq3rdorder}
\end{align}
These equations obviously suggest the following two different cases: $V_4=0$ or $V_4\neq 0$.
\subsection*{Case 1. $V_4=0$}
In this case, all the determining equations associated with the third-order terms vanish. Therefore we go on to investigate the determining equations coming from the second-order terms. From these terms, we find the following determining equations:
\begin{align}
    & rf_{10}(2V_1-6\hbar V_5+\hbar r V'_5)+\hbar(f_{10}'+(f_5+f_6)V_5')=0, \label{seq1}\\
    & r^3 f_{10}(2V_1-6\hbar V_5+\hbar r V_5')+\hbar(r^2 f_{10}'+f_2'+f_7'+r^2(f_2+f_5+f_6+f_7)V_5')=0, \label{seq2}\\
    & \hbar( f_2'+f_7')+r (f_5+f_6)(-V_1+3\hbar V_5+\hbar r V_5')+f_{10}(-\hbar r+r^3(2V_1-6\hbar V_5+\hbar r V_5'))=0, \label{seq3}\\
    & f_8'+f_9'=0, \label{seq4}\\
    & \hbar (r^2f_{10}+f_2+f_7)+r^2(f_5+f_6)(V_1-3\hbar V_5)=0, \label{seq5}\\
    &\hbar f_{10}+(f_5+f_6)(V_1-3\hbar V_5)+\hbar r (f_2+f_7)V_5'=0, \label{seq7}\\
    & r(f_5+f_6)(-V_1+3\hbar V_5+\hbar rV_5')-\hbar f_6'+\mathrm{i}\hbar f_8'+f_{10}(r^3V_1+\hbar r(-2-3r^2V_5+r^3V_5'))=0, \label{seq8}\\
    & r(f_5+f_6)(-V_1+3\hbar V_5+\hbar rV_5')-\hbar f_5'+\mathrm{i}\hbar f_9'+f_{10}(r^3V_1+\hbar r(-2-3r^2V_5+r^3V_5'))=0, \label{seq9}\\
    & 2r(f_5+f_6)(-V_1+3\hbar V_5+\hbar rV_5')-\hbar (f_5'+f_6') \nonumber\\ &\qquad \qquad +2f_{10}(r^3V_1+\hbar r(-2-3r^2V_5+r^3V_5'))=0,  
    \label{seq10}\\
    & r(f_5+f_6)(-V_1+3\hbar V_5+2\hbar rV_5')-\hbar (f_5'+f_6') \nonumber\\ &\qquad \qquad +\hbar r^2(f_2+f_7)V_5' +f_{10}(2r^3V_1+\hbar r(-3-6r^2V_5+2r^3V_5'))=0,  \label{seq11}\\
    & \hbar r^3(r^2 f_{10}+f_2+f_7)V_5'+(f_5+f_6)(\hbar-r^2 V_1+3\hbar r^2 V_5+\hbar r^3 V_5')=0, \label{seq12}\\
    & r^3f_{10}(2V_1-6\hbar V_5+\hbar rV_5')+\hbar(r^2f_{10}'-f_2'+f_7')+\hbar r^2(f_2+f_5+f_6-f_7)V_5'=0, \label{seq13}\\
    & \hbar(r^2f_{10}-f_2+f_7)+r^2(f_5+f_6)(V_1-3\hbar V_5)+2\hbar r^3 f_7 V_5'=0, \label{seq14} \\
    & -\hbar(f_5'+f_6'+2f_7'-2r^2(f_5+f_6+f_7)V_5')+2f_{10}(r^3V_1+\hbar r(-1-3r^2V_5+r^3V_5'))=0, \label{seq15} \\
    & \hbar r^3(r^2f_{10}-f_2+f_7)V_5'+(f_5+f_6)(\hbar-r^2V_1+3\hbar r^2V_5+\hbar r^3V_5')+2\hbar f_7=0. \label{seq16} 
\end{align}
From equations \eqref{seq8} and \eqref{seq9}, we directly see that $f_8$ and $f_9$ must be constants, and using these equations again we obtain $-f_5'+f_6'=0$. In addition from \eqref{seq2}, \eqref{seq5}, \eqref{seq13}, and \eqref{seq14}, we find that $f_2+rf_2'=0$. Therefore we get
\begin{equation}\label{rel1}
    f_9=c_1,\quad f_8=c_2,\quad f_6=f_5+c_3,\quad f_2=\dfrac{c_4}{r},
\end{equation}
where $c_i$ ($i=1,2,3,4$) are real constants. Calculating $2r(\eqref{seq2}-r^2 \eqref{seq1})+\eqref{seq16}-\eqref{seq12}+\eqref{seq5}-\eqref{seq14}$ and introducing the relations in \eqref{rel1} into the result give $f_7+r f_7'=0$. Solving this differential equation for $f_7$ yields $f_7=c_5/r$, where $c_5$ is a real constant.
Now we introduce this relation together with the relations in \eqref{rel1} into the determining equations listed above. Notice that from these relations \eqref{seq5}, \eqref{seq7}, \eqref{seq12}, \eqref{seq14}, and \eqref{seq16} respectively become
\begin{align}
    & \hbar r^2 f_{10}+r^2(c_3+2f_5)(V_1-3\hbar V_5)+\dfrac{\hbar(c_4+c_5)}{r}=0, \label{seq5n}\\
    & \hbar f_{10}+(c_3+2f_5)(V_1-3\hbar V_5)+(c_4+c_5)\hbar V_5'=0, \label{seq7n}\\
    & \hbar r^2(c_4+c_5+r^3f_{10})V_5'+(c_3+2f_5)(\hbar-r^2 V_1+3\hbar r^2V_5+\hbar r^3 V_5')=0, \label{seq12n}\\
    & \hbar r^2 f_{10}+r^2(c_3+2f_5)(V_1-3\hbar V_5)+\frac{\hbar}{r}(-c_4+c_5+2c_5r^3V_5')=0, \label{seq14n} \\
    & (c_3+2f_5)(\hbar-r^2V_1+3\hbar r^2V_5+\hbar r^3V_5')+\frac{\hbar}{r}(2c_5+r^3(-c_4+c_5+r^3f_{10})V_5')=0. \label{seq16n}
\end{align}
It is easy to see that considering $\eqref{seq5n}-r^2 \eqref{seq7n}$ yields $(c_4+c_5)(-1+r^3V_5')=0$. Therefore we consider the following two subcases: $c_5=-c_4$ or $-1+r^3V_5'=0$, i.e., $V_5=-\dfrac{1}{2r^2}+\alpha_1$, where $\alpha_1$ is a real constant. 
Before moving on to the analysis of these two subcases, we also present the determining equations coming from the first- and zeroth- order terms. Again we will not write down the complete set of these equations, instead we will proceed by solving some of these equations step by step. Introducing the above relations, we have the following equations:
\begin{align}
    &\mathrm{i}(-2r^3(c_1+c_2)V_3+2r^3 f_4(V_1-3\hbar V_5)+\hbar r^2 f_4')+\hbar r^2(c_3+2f_5)(V_1'+\hbar V_5')\nonumber \\
    &\qquad\qquad +\hbar r^2(7\hbar+r^2 V_1-3\hbar r^2 V_5)f_{10}'+\hbar(-\hbar+2r^2V_1-6\hbar r^2 V_5)f_5'+\hbar^2 r(r^2 f_{10}''+f_5'') \nonumber \\
    &\qquad\qquad+\hbar r^3(10V_1-30\hbar V_5+rV_1'+\hbar r V_5')f_{10}=0, \label{feq1}\\
      &-\mathrm{i}(2r^5(c_1+c_2)V_3-2r^5 f_4(V_1-3\hbar V_5)-\hbar r^2 (f_1'+f_3'+r^2f_4'))+\hbar r^2(c_3+2f_5)(V_1'+\hbar V_5')\nonumber \\
    &\qquad\qquad +\hbar r^4(7\hbar+r^2V_1-3\hbar r^2 V_5)f_{10}'+2\hbar r^4(V_1-3\hbar V_5)f_5'+\hbar^2 r^3(r^2 f_{10}''+f_5'') \nonumber \\
    &\qquad\qquad+\hbar r^3(\hbar+9r^2V_1-27\hbar r^2 V_5+r^3V_1')f_{10}+\hbar^2(c_4+c_5)(-2+r^3V_5')=0, \label{feq2}\\
    & r(c_3+2f_5)(2r V_3+\hbar^2 V_5')+\hbar(c_4+c_5)V_1'+\hbar^2 r^3 f_{10}V_5'=0, \label{feq3} \\
     &-\mathrm{i}r^2(2rf_4+r^2f_4'+f_1'+f_3')+2\hbar (c_4+c_5)+r^3 f_{10}(-8\hbar-2r^2V_1+6\hbar r^2 V_5)\nonumber \\
    &\qquad\qquad -\hbar r^2(8r^2 f_{10}'+6f_5'+r^3f_{10}''+2r f_5'')=0, \label{feq4}\\
    &-\mathrm{i}r( (c_1-c_2)(\hbar V_1-4V_2-\hbar^2 V_5)+4c_2 r^2V_3+2f_4(\hbar-r^2 V_1+3\hbar r^2 V_5) ) \nonumber \\
    &\qquad\qquad +2r^2(2r V_3+\hbar(V_1'+\hbar V_5'))f_5+\hbar r(-7\hbar+7r^2 V_1-21\hbar r^2 V_5+r^3 V_1'+\hbar r^3 V_5')f_{10} \nonumber \\
    &\qquad\qquad +2 \hbar (r^2 V_1-3\hbar(1+r^2 V_5))f_5'-\hbar r^2(\hbar-r^2(V_1-3\hbar V_5))f_{10}'+c_4 \hbar r(V_1'+\hbar V_5') \nonumber \\
    &\qquad\qquad +c_5\hbar^2 r V_5'-\hbar^2 f_5''+c_3 r(-\hbar V_1+4V_2+4r^2 V_3+\hbar^2 V_5+\hbar r(V_1'+\hbar V_5'))=0, \label{feq5}\\
    &-2 c_3 \hbar r (-\hbar V_1+4V_2+2r^2 V_3+\hbar^2 V_5)=0 \label{feq9}\\
    &-\mathrm{i}(2r^5(c_1+c_2)V_3-2r^5 f_4(V_1-3\hbar V_5)-\hbar r^2 (-f_1'+f_3'+r^2f_4'))+\hbar r^4(c_3+2f_5)V_1'\nonumber \\
    &\qquad\quad +\hbar r^4(7\hbar+r^2V_1-3\hbar r^2 V_5)f_{10}'+2\hbar r^4(V_1-3\hbar V_5)f_5'+\hbar^2 r^3(r^2 f_{10}''+f_5'') \nonumber \\
     &\qquad\quad+\hbar r^3(\hbar+9r^2V_1-27\hbar r^2 V_5+r^3V_1')f_{10}+\hbar^2(c_4-c_5)(2+r^3V_5')-2\hbar^2 c_5 r^3 V_5'=0, \label{feq10}\\
     & 2\mathrm{i}r^2 (6r(c_1+c_2)V_3-6rf_4(V_1-3\hbar V_5)-\hbar(6f_4'+rf_4''))+6\hbar r^2(-6\hbar-r^2 V_1+3\hbar r^2 V_5)f_{10}' \nonumber \\
     &\qquad\quad -2r^3(15\hbar(V_1-3\hbar V_5)+2r(V_0'+V_2'))f_{10}-\hbar^2 r(13 r^2 f_{10}''+8f_5'')-\hbar^2 r^2(r^2 f_{10}'''+2f_5''') \nonumber \\
     &\qquad\quad +8r^2(rV_3-V_0'-V_2')f_5+4\hbar(2\hbar-3r^2 V_1+9 r^2 V_5)f_5' \nonumber \\
     &\qquad\quad -4r^2 ((2(c_4+c_5)-c_3 r)V_3+c_3(V_0'+V_2')+(c_4+c_5)rV_3')=0, \label{zeq1} \\
     &2\mathrm{i}\left( 4(c_1+c_2)r^3 V_3-2r f_4(\hbar+2r^2 V_1-6\hbar r^2 V_5)+\hbar(2f_1'-2f_3'-r(6rf_4'-f_1''+f_3''+r^2 f_4'')) \right) \nonumber\\
     &\qquad\quad +8r^2(4r V_3-V_0'+V_2'+r^2 V_3')f_5-8 \hbar r^2(V_1-3\hbar V_5)f_5'-10\hbar^2 rf_5''-2\hbar^2 r^2 f_5''' \nonumber\\
     &\qquad\quad +(-10\hbar r(\hbar+r^2 V_1)+8r^5 V_3+60\hbar^2 r^3 V_5+4r^4(-V_0'+V_2'+r^2 V_3') )f_{10} \nonumber \\
     &\qquad\quad -2\hbar r^2(19\hbar+2r^2 V_1-6\hbar r^2 V_5 )f_{10}'-13\hbar^2 r^3 f_{10}''-\hbar^2 r^4 f_{10}'''  \nonumber \\
     &\qquad\quad +4r(-c_4+c_5+c_3 r)(2r V_3+r^2 V_3'-V_0'+V_2')+8r(2c_5 V_2'+c_3r^2 V_3)=0. \label{zeq2}
\end{align}
Note that from \eqref{feq2} and \eqref{feq10}, $f_1$ is constant. From \eqref{feq4} we directly have $2rf_4+r^2f_4'+f_1'+f_3'=0$. Then from \eqref{feq1} and \eqref{feq2} $f_3$ is also constant. Thus we have
\begin{eqnarray}
    f_4=\dfrac{c_7-c_8-c_9}{r^2},\quad f_1=c_8,\quad f_3=c_9,
\end{eqnarray}
where $c_7,c_8,c_9$ are real constants. Note also that even though equations \eqref{feq1}--\eqref{zeq2} are given in complex form, we should of course consider that their real and imaginary parts are simultaneously equal to zero since all $V_i$'s and $f_j$'s are real functions of $r$. However, for the sake of simplicity, we prefer to write these equations in this compact form and avoid writing much more equations. Now let us go back to the two subcases:
\subsubsection*{Subcase 1. $V_5=-\dfrac{1}{2r^2}+\alpha_1$}
Introducing this potential into \eqref{seq5n} and \eqref{seq14n} and subtracting the resulting equations, we directly obtain $c_5=c_4$. Then \eqref{seq12n} and \eqref{seq14n} become
\begin{align}
    & \dfrac{\hbar(2c_4+r^3 f_{10})}{r}+(c_3+2f_5)\left( \dfrac{\hbar}{2}+3\alpha_1 \hbar r^2-r^2 V_1\right)=0, \label{seq12nn}\\
    & \dfrac{\hbar(2c_4+r^3 f_{10})}{r}+(c_3+2f_5)\left( \dfrac{3\hbar}{2}-3\alpha_1 \hbar r^2+r^2 V_1\right)=0. \label{seq14nn}
\end{align}
Summing up these two equations we get 
\begin{equation}
    f_{10}=-\dfrac{2c_4+c_3 r+2r f_5}{r^3}.
\end{equation}
Subtracting \eqref{seq12nn} from \eqref{seq14nn}, we find that 
\begin{equation}
    (c_3+2f_5)(-\hbar +6\alpha_1 \hbar r^2-2r^2 V_1)=0,
\end{equation}
which yields the following cases: $f_5=-c_3/2$ or $-\hbar +6\alpha_1 \hbar r^2-2r^2 V_1=0$, i.e., $V_1=\dfrac{-\hbar}{2r^2}+3\alpha_1\hbar$. Now we continue our analysis by investigating these cases separately.
\paragraph{S1. $V_1=\dfrac{-\hbar}{2r^2}+3\alpha_1\hbar$.} Substituting all the above relations into the remaining determining equations derived from the second-order terms gives us $f_5=c_6$, where $c_6$ is a real constant. In this case, all determining equations coming from the second-order terms vanish. Now we analyze the determining equations coming from the first-order and the zeroth-order terms. Using \eqref{feq1} and \eqref{feq3}, we find that
\begin{align*}
    & (c_1+c_2)V_3=0, \\
    & (c_3+2c_6)V_3=0.
\end{align*}
We proceed by exploring the following possibilities: $V_3=0$ and $V_3\neq 0$.
\paragraph{I. $V_3=0$.} Introducing these relations, \eqref{feq5} reduces to
\[ (c_1-c_2-\mathrm{i}c_3)(\alpha_1 \hbar^2-2 V_2)=0. \]
This equation leads us to the following two cases.
\paragraph{A. $V_2=\alpha_1 \hbar^2/2$.} Then from \eqref{zeq1} we have $c_4 V_0'=0$. Then again we need to consider two situations.

The first one is $V_0=\alpha_2$, where $\alpha_2$ is a real constant. In this case, all the determining equations are satisfied and we have eight arbitrary constants. The integrals of motion corresponding to $c_1, c_2, c_3, c_4, c_6,$ and $c_7$ are given as follows. 
\begin{align}
    &Y_s^1=(\vec{\sigma}_2, \vec{L}), \label{IoM1}\\
    & Y_s^2=(\vec{\sigma}_1, \vec{L}), \label{IoM2} \\
    & Y_s^3=\dfrac{1}{r^2}(\vec{\sigma}_1,\vec{x})(\vec{\sigma}_2,\vec{x})\left(-(\vec{x},\vec{p})+\dfrac{3\mathrm{i}\hbar}{2}\right)+(\vec{\sigma}_2,x)(\vec{\sigma}_1,p), \label{IoM3}\\
    & Y_s^4=-\dfrac{\mathrm{i}\hbar}{r}+\dfrac{1}{r}(1+(\vec{\sigma}_1, \vec{\sigma}_2))(\vec{x},\vec{p})+\dfrac{2}{r^3}(\vec{\sigma}_1,\vec{x})(\vec{\sigma}_2,\vec{x})(-(\vec{x},\vec{p})+\mathrm{i}\hbar)-\dfrac{\mathrm{i}\hbar}{2r}(\vec{\sigma}_1, \vec{\sigma}_2), \label{IoM4}\\
    & Y_s^5=(\vec{\sigma}_1, \vec{x})(\vec{\sigma}_2, \vec{p})+(\vec{\sigma}_2, \vec{x})(\vec{\sigma}_1, \vec{p})+\dfrac{1}{r^2}(\vec{\sigma}_1,\vec{x})(\vec{\sigma}_2,\vec{x})(-2(\vec{x},\vec{p})+3\mathrm{i}\hbar), \label{IoM5}\\
    & Y_s^6=\dfrac{(\vec{\sigma}_1,\vec{x})(\vec{\sigma}_2,\vec{x})}{r^2}. \label{IoM6}
\end{align}
The integrals of motion corresponding to $c_8$ and $c_9$ are respectively $1-Y_s^6$ and $(\vec{\sigma}_1, \vec{\sigma}_2)-Y_s^6$. Note that $1$ and $(\vec{\sigma}_1, \vec{\sigma}_2)$ are trivial integrals of motion, that is, they are integrals of motion for arbitrary potentials in the Hamiltonian. Therefore, we can ignore these two integrals. 

The second situation is $c_4=0$. In this case, all the determining equations are satisfied and $V_0$ is an arbitrary potential. We have the same integrals of motion as listed above, but $Y_s^4$.
\paragraph{B. $c_2=c_1, c_3=0$.} Then from \eqref{zeq1} 
\begin{equation}
    c_4(V_0'+V_2')=0.
\end{equation}
This results in the following classifications: $V_0'+V_2'=0$ or $c_4=0$.

Firstly suppose that $V_2=-V_0+\alpha_3$, where $\alpha_3$ is a real constant. So all the determining equations are satisfied. There are six arbitrary constants. The integrals of motion for $c_4$, $c_6$, and $c_7$ are $Y_s^4$, $Y_s^5$, and $Y_s^6$. The integrals of motion for $c_8$ and $c_9$ can be ignored, since they reduce to $Y_s^6$. The integral of motion for $c_1$ is
\begin{equation}
    Y_s^7=(\vec{\sigma}_1+\vec{\sigma}_2, \vec{L}). \label{IoM7}
\end{equation}

Now suppose that $c_4=0$. Now all the determining equations are satisfied for arbitrary $V_0$ and $V_2$. We have five integrals of motion that correspond to five arbitrary constants. These are the same as the ones in the previous case but $Y_s^4$.
\paragraph{II. $V_3\neq 0$.} In this case we have $c_2=-c_1$ and $c_6=-c_3/2$. If we substitute these relations into \eqref{feq5}, we obtain 
\[ (2\mathrm{i}c_1+c_3)(-\alpha_1 \hbar^2+2 V_2+r^2 V_3)=0, \]
 which yields two cases.
 \paragraph{A. $-\alpha_1 \hbar^2+2 V_2+r^2 V_3= 0$.} This gives us 
 \[ V_3= \dfrac{\alpha_1 \hbar^2-2V_2}{r^2}.\]
 Then using \eqref{zeq1} we find that $c_4(V_0'+3V_2')=0$. This leads us to explore two distinct paths.

 If $V_0'+3V_2'=0$, then $V_2=-\frac{V_0}{3}+\alpha_4$, where $\alpha_4$ is a real constant. Now all the determining equations are satisfied for an arbitrary potential $V_0$. There are six arbitrary constants, $c_1,c_3, c_4, c_7, c_8,$ and $c_9$. The integrals of motion for $c_1$ and $c_3$ are
 \begin{align}
      & Y_s^8=(\vec{\sigma}_1-\vec{\sigma}_2, \vec{L}), \label{IoM8} \\
      & Y_s^9=(\vec{\sigma}_1, \vec{x})(\vec{\sigma}_2, \vec{p})-(\vec{\sigma}_2, \vec{x})(\vec{\sigma}_1, \vec{p}). \label{IoM9}
 \end{align}
 The integrals of motion for $c_4$ and $c_7$ are $Y_s^4$ and $Y_s^6$, and the integrals of motion for $c_8$ and $c_9$ are the trivial ones.

If $c_4=0$, then all the determining equations are satisfied for arbitrary $V_0$ and $V_2$. We have five integrals of motion corresponding to five arbitrary constants $c_1,c_3, c_7, c_8,$ and $c_9$. These are $Y_s^6, Y_s^8, Y_s^9$, and two trivial integrals of motion.
\paragraph{B. $c_3=0, c_1=0$.} Then all the determining equations obtained by the first-order terms would be zero. Introducing these equalities into \eqref{zeq1}, we have $c_4(2rV_3+r^2V_3'-V_0'-V_2')=0$. We have two cases to consider.

The first case is $2rV_3+r^2V_3'-V_0'-V_2'$. We solve this equation for $V_3$ to obtain $V_3=(V_0+V_2+\alpha_5)/r^2$, where $\alpha_5$ is a real constant. Then all the determining equations are satisfied. We have four arbitrary constants, but only two of them give non-trivial integrals of motion. For $c_4$ we have $Y_s^4$ and for $c_7$ we have $Y_s^6$.

The second case is $c_4=0$. In this case the full set of determining equations are satisfied for arbitrary potentials $V_0, V_2,$ and $V_3$. Only one non-trivial integral of motion that comes from the arbitrary constant $c_7$ is $Y_s^6$. 
\paragraph{S2. $f_5=-c_3/2$.} From \eqref{feq9} we have
\begin{equation}
    c_3\left( \hbar^2\left( \alpha_1-\dfrac{1}{2r^2} \right)-\hbar V_1+4V_2+2r^2 V_3 \right)=0. \label{f2eq9}
\end{equation}
This equation suggests two possibilities for analysis.
\paragraph{I. $c_3\neq0$.} In this case from \eqref{f2eq9} we have $\hbar^2\left( \alpha_1-\dfrac{1}{2r^2} \right)-\hbar V_1+4V_2+2r^2 V_3 =0$. Solving this equation for $V_3$, we find that
\[ V_3=\dfrac{\hbar^2}{4r^4}-\dfrac{\alpha_1 \hbar^2-\hbar V_1+4V_2}{2r^2}. \]
Substituting this relation together with the previous relations into \eqref{feq5}, we obtain
\begin{align}
   &\dfrac{\hbar(-\hbar(c_1+c_2)(1-2\alpha_1 r^2)+2(c_7-c_8-c_9)(1-6\alpha_1 r^2))}{2r} \nonumber\\
   &\qquad\quad-(2(-c_7+c_8+c_9)+\hbar(c_1+c_2))r V_1+4(c_1+c_2)r V_2=0. \label{f3eq5}
\end{align}
Based on this equation, we examine the following situations.
\paragraph{A. $c_1+c_2\neq0$.} We can solve \eqref{f3eq5} for $V_2$ to get
\begin{align}
    V_2=&\dfrac{-\hbar(-\hbar(c_1+c_2)(1-2\alpha_1 r^2)+2(c_7-c_8-c_9)(1-6\alpha_1 r^2))}{8(c_1+c_2)r^2}\nonumber\\
    & +\dfrac{(2(-c_7+c_8+c_9)+\hbar(c_1+c_2))r V_1}{4(c_1+c_2)r}
\end{align}
Now all the determining equations are satisfied for arbitrary $V_0$ and $V_1$. The integral of motion depends on $c_1, c_2, c_3, c_7, c_8$, and $c_9$. However the potentials $V_2$ and $V_3$ also depend on $\beta=(-c_7+c_8+c_9)/(c_1+c_2)$. Hence, we can choose $c_9=c_7-c_8+\beta(c_1+c_2)$. Therefore we have the following non-trivial integrals of motion for this case: $Y_s^9$,
\begin{align}
    &Y_s^{10}=(\vec{\sigma}_2, \vec{L})-\dfrac{\beta}{r^2}(\vec{\sigma}_1,\vec{x})(\vec{\sigma}_2,\vec{x}), \label{IoM10} \\
     &Y_s^{11}=(\vec{\sigma}_1, \vec{L})-\dfrac{\beta}{r^2}(\vec{\sigma}_1,\vec{x})(\vec{\sigma}_2,\vec{x}). \label{IoM11}
\end{align}
\paragraph{B. $c_1+c_2=0$.} Putting $c_2=-c_1$ in the remaining determining equations 
\[ (c_7-c_8-c_9)(-\hbar+6\hbar \alpha_1 r^2-2r^2 V_1)=0. \]
Then we have two cases. The first one is 
\[ V_1=-\dfrac{\hbar}{2r^2}+3\hbar\alpha_1. \]
However we will omit this case from our analysis, as it has already been thoroughly investigated. Therefore we set $c_9=c_7-c_8$. Then all the determining equations are satisfied for arbitrary $V_0, V_1,$ and $V_2$. There are four arbitrary constants $c_1,c_3,c_7$, and $c_8$. The integrals of motion corresponding to $c_7$ and $c_{8}$ are trivial. We have two non-trivial integrals of motion $Y_s^8$ and $Y_s^9$. 
\paragraph{II. $c_3=0$.} From \eqref{feq1}, we get 
\[ (c_7-c_8-c_9)(\hbar(-1+6\alpha_1 r^2)-2r^2 V_1 )+2(c_1+c_2)r^4 V_3=0.\]
This equation gives rise to the following circumstances.
\paragraph{A. $c_1+c_2\neq 0$.} In this case we directly have
\[ V_3=\dfrac{(-c_7+c_8+c_9)(\hbar(-1+6\alpha_1 r^2)-2r^2 V_1)}{2(c_1+c_2)r^4}. \]
Introducing this relation into \eqref{feq5} 
\begin{align*}
    &(c_1-c_2)( 2(c_7-c_8-c_9)(\hbar-6\hbar \alpha_1 r^2+2r^2 V_1)-\hbar(c_1+c_2)(\hbar-2\hbar \alpha_1 r^2+2r^2 V_1) \nonumber \\
    &\qquad\qquad +8(c_1+c_2)r^2 V_2 )=0.
\end{align*}
We need to consider two possibilities. The first one is 
\[ V_2=-\dfrac{2(c_7-c_8-c_9)(\hbar-6\hbar \alpha_1 r^2+2r^2 V_1)-\hbar(c_1+c_2)(\hbar-2\hbar \alpha_1 r^2+2r^2 V_1)}{8(c_1+c_2)r^2}. \]
Then all the determining equations are satisfied for arbitrary $V_0$ and $V_1$. Note that this case is nothing but the case S2-I-A with $c_3=0$. Therefore we have only $Y_s^{10}$ and $Y_s^{11}$ as non-trivial integrals of motion.

Now let us assume that $c_2=c_1$. All the determining equations are satisfied for arbitrary $V_0, V_1,$ and $V_2$. We have $4$ arbitrary constants, $c_1, c_7, c_8,$ and $c_9$. However the potential $V_3$ depends on $\beta=(-c_7+c_8+c_9)/(2c_1)$. So we can choose $c_9=c_7-c_8+2\beta c_1$. We have only one non-trivial integral of motion:
\begin{equation}
    Y_s^{12}=(\vec{\sigma}_1+\vec{\sigma}_2, \vec{L})-\dfrac{2\beta}{r^2}(\vec{\sigma}_1, \vec{x})(\vec{\sigma}_2, \vec{x}). \label{IoM12}
\end{equation}
\paragraph{B. $c_1+c_2=0$.} From \eqref{feq1}, we obtain
\[ (c_7-c_8-c_9)(-\hbar+6\hbar\alpha_1 r^2-2r^2 V_1)=0. \]
The analysis naturally splits into two branches.
\paragraph{a. $-\hbar+6\hbar\alpha_1 r^2-2r^2 V_1=0$.} Solving this equation for $V_1$ gives
\[ V_1=-\dfrac{\hbar}{2r^2}+3\hbar \alpha_1.\]
Now using \eqref{feq5} we get
\[ c_1(-\alpha_1\hbar^2+2V_2+r^2 V_3)=0. \]
There are two options to consider. Suppose that $V_3=\dfrac{\alpha_1 \hbar^2-2 V_2}{r^2}$. Then all the determining equations are satisfied for arbitrary $V_0$ and $V_2$. We have four arbitrary constants, but there are only two non-trivial integrals of motion: $Y_s^6$ and $Y_s^8$. 

Now suppose that $c_1=0$. All the determining equations are satisfied for arbitrary $V_0, V_2,$ and $V_3$, and we have only one non-trivial integrals of motion that is $Y_s^6$.
\paragraph{b. $c_7-c_8-c_9=0$.} We immediately have $c_9=c_7-c_8$, which is introduced in \eqref{feq5} to give
\[ c_1(-\hbar^2+2\alpha_1 \hbar^2 r^2-2\hbar r^2 V_1+8r^2 V_2+4r^4 V_3)=0. \]
This equation yields two cases. Suppose that 
\[ V_3=\dfrac{\hbar^2-2\alpha_1 \hbar^2 r^2+2\hbar r^2 V_1-8r^2 V_2}{4r^4}. \]
Then all the determining equations are satisfied for arbitrary $V_0, V_1,$ and $V_2$. There is only one non-trivial integral of motion, $Y_s^8$.

Now suppose that $c_1=0$. In this case all the determining equations are satisfied for arbitrary $V_0, V_1, V_2,$ and $V_3$, however, there is no non-trivial integral of motion.

This means that we have completed the analysis of Subcase 1. 
\subsubsection*{Subcase 2. $c_5=-c_4$}
Subtracting \eqref{seq12n} from \eqref{seq16n}, we find that 
\[ c_4(1+r^3 V_5')=0. \]
Hence we need to consider the following cases: $c_4=0$ or $1+r^3 V_5'=0$, i.e., $V_5=\dfrac{1}{2r^2}+\alpha_6$, where $\alpha_6$ is a real constant. 
\paragraph{S1. $V_5=\dfrac{1}{2r^2}+\alpha_6$.} Introducing these relations, \eqref{seq13} and \eqref{seq15} become
\begin{align}
    & f_{10}(-2 r^2(2\hbar+3\alpha_6\hbar r^2-r^2 V_1))-\hbar(c_3+2f_5-r^3 f_{10}')=0, \label{seq13n}\\
    & f_{10}(-r^2(7\hbar+6\alpha_6\hbar r^2-2r^2 V_1))-2\hbar(c_3+2f_5+r f_{5}')=0. \label{seq15n}
\end{align}
From these equations we find that
\[ f_{10}=\dfrac{c_{11}-c_3 r-2r f_5}{r^3}, \]
where $c_{11}$ is a real constant. Using \eqref{seq16n} we obtain
\begin{equation}\label{seq16nn}
    \dfrac{5c_3\hbar}{2}-\dfrac{c_{11}\hbar}{r}+3\alpha_6 c_3 \hbar r^2-c_3 r^2 V_1+f_5(5\hbar+ 6\alpha_6 \hbar r^2-2r^2 V_1)=0.
\end{equation}
Our analysis continues based on the following two alternatives.
\paragraph{I. $5\hbar+ 6\alpha_6 \hbar r^2-2r^2 V_1\neq 0$.} In this case, solving \eqref{seq16nn} for $f_5$
\[ f_5=\dfrac{2c_{11}\hbar-5c_3\hbar r-6\alpha_6 c_3\hbar r^3+2c_3 r^3 V_1}{2r(5\hbar+ 6\alpha_6 \hbar r^2-2r^2 V_1)}. \]
We introduce these relations into the remaining determining equations coming from the second-order terms and obtain
\begin{align}
   & c_{11}( \hbar^3(115+414\alpha_6 r^2 +540\alpha_6^2 r^4+216\alpha_6^3 r^6 )-6\hbar^2 r^2(23+60\alpha_6r^2+36\alpha_6^2 r^4)V_1 \nonumber\\
    &\qquad +12 \hbar r^4(5+6\alpha_6 r^2)V_1^2 -8r^6 V_1^3+4\hbar^2 r^3 V_1')=0. \label{sc2s1IA-1}
\end{align}
This equation leads to the consideration of two situations.
\paragraph{A. $c_{11}\neq 0$.} In this case we can solve the nonlinear differential equation given in the second factor of the expression on the left-hand side of \eqref{sc2s1IA-1}. The solution is
\[ V_1=\dfrac{5\hbar}{2r^2}+3\hbar \alpha_6+\dfrac{\epsilon\hbar}{r^2\sqrt{2+\alpha_7\hbar^2 r^2}}, \]
where $\alpha_7$ is a real constant and $\epsilon^2=1$. Then all the determining equations obtained by the second-order terms would be zero. We will proceed by solving the determining equations coming from the first-order and zeroth-order terms. Using \eqref{feq3} and from the fact that $c_{11}\neq 0$, we get
\[ V_3=-\dfrac{\epsilon\hbar^2}{2r^4\sqrt{2+\alpha_7\hbar^2 r^2}}.\]
Then from \eqref{feq1}, we find that $c_9=c_7-c_8+\hbar(c_1+c_2)/2$. Introducing these relations into the remaining determining equations yields
\[ (c_1-c_2-\mathrm{i} c_3)(\hbar^2+\epsilon \hbar^2(1+\alpha_6 r^2)\sqrt{2+\alpha_7\hbar^2 r^2}-2\epsilon r^2 \sqrt{2+\alpha_7\hbar^2 r^2}V_2)=0. \]
Hence we consider the following two cases.

\paragraph{a.} Suppose that 
\[ V_2= \dfrac{\hbar^2}{2r^2}+\dfrac{\epsilon \hbar^2}{2r^2 \sqrt{2+\alpha_7\hbar^2 r^2}}+\dfrac{\hbar^2\alpha_6}{2}.  \]
Now from \eqref{zeq1}, since $c_{11}\neq 0$ we obtain
\[ V_0'+\dfrac{3\hbar^2}{r^3}+\dfrac{\epsilon\hbar^2(4+3\alpha_7 \hbar^2 r^2)}{r^3(2+\alpha_7\hbar^2 r^2)^{3/2}}=0.\]
Solving this differential equation for $V_0$
\[ V_0=\dfrac{3\hbar^2}{2r^2}+\dfrac{\epsilon\hbar^2}{r^2 \sqrt{2+\alpha_7\hbar^2 r^2}}+\alpha_8,\]
where $\alpha_8$ is a real constant. So all the determining equations are satisfied. We have seven arbitrary constants, two of which $c_7$ and $c_8$ give trivial integrals of motion. The non-trivial integrals of motion are as follows: $Y_s^9$,
\begin{align}
       &Y_s^{13}=(\vec{\sigma}_2, \vec{L})-\dfrac{\hbar}{2r^2}(\vec{\sigma}_1,\vec{x})(\vec{\sigma}_2,\vec{x}), \label{IoM13} \\
     &Y_s^{14}=(\vec{\sigma}_1, \vec{L})-\dfrac{\hbar}{2r^2}(\vec{\sigma}_1,\vec{x})(\vec{\sigma}_2,\vec{x}), \label{IoM14}\\
     &Y_s^{15}=-\dfrac{\mathrm{i}\hbar}{r}+\dfrac{1}{r}(1-(\vec{\sigma}_1,\vec{\sigma}_2))(\vec{x}, \vec{p})+\dfrac{\mathrm{i}\hbar}{r}(\vec{\sigma}_1,\vec{\sigma}_2), \label{IoM15}\\
     &Y_s^{16}=\left( \dfrac{1+\epsilon \sqrt{2+\alpha_7 \hbar^2 r^2}}{r^3} \right)(\vec{\sigma}_1,\vec{x})(\vec{\sigma}_2,\vec{x})(\vec{x}, \vec{p})+\dfrac{\epsilon \mathrm{i}\hbar}{2r}\sqrt{2+\alpha_7 \hbar^2 r^2} (\vec{\sigma}_1, \vec{\sigma}_2) \nonumber \\
     &\qquad\quad -\dfrac{\epsilon \mathrm{i}\hbar(4+2r+2\epsilon \sqrt{2+\alpha_7 \hbar^2 r^2}+3\alpha_7\hbar^2 r^2) }{2r^3 \sqrt{2+\alpha_7 \hbar^2 r^2}}(\vec{\sigma}_1,\vec{x})(\vec{\sigma}_2,\vec{x}) \nonumber \\
     &\qquad \quad -\dfrac{\epsilon}{2r}\sqrt{2+\alpha_7 \hbar^2 r^2}\left((\vec{\sigma}_1,\vec{x})(\vec{\sigma}_2,\vec{p})+(\vec{\sigma}_2,\vec{x})(\vec{\sigma}_1,\vec{p})\right) \label{IoM16}.
\end{align}

\paragraph{b.} Now suppose that $c_2=c_1$ and $c_3=0$. From these relations, we see that the only non-vanishing determining equations are \eqref{zeq1} and \eqref{zeq2}. By \eqref{zeq1}
 \[ 2(V_0'+V_2')+\dfrac{3\epsilon \hbar^2(4+3\alpha_7 \hbar^2 r^2)}{r^3(2+\alpha_7\hbar^2 r^2)^{3/2}}+\dfrac{8\hbar^2}{r^3}=0.\]
From this equation we find that
\[ V_2=\dfrac{2\hbar^2}{r^2}+\dfrac{3\epsilon \hbar^2}{2r^2 \sqrt{2+\alpha_7\hbar^2 r^2}}-V_0 +\alpha_9, \]
where $\alpha_9$ is a real constant. Then from \eqref{zeq2} 
\begin{align*}
    (c_{11}-4c_4)\left( \epsilon\hbar^2(4+3\alpha_7\hbar^2 r^2)+(2+\alpha_7\hbar^2 r^2)^{3/2} (3\hbar^2+r^3 V_0') \right)=0.
\end{align*}
We proceed by analyzing the following cases.

The first case is $\epsilon\hbar^2(4+3\alpha_7\hbar^2 r^2)+(2+\alpha_7\hbar^2 r^2)^{3/2} (3\hbar^2+r^3 V_0') =0$. Solving this equation for $V_0$, we obtain
\[ V_0=\dfrac{3\hbar^2}{2r^2}+\dfrac{\epsilon \hbar^2}{r^2 \sqrt{2+\alpha_7\hbar^2 r^2}}+\alpha_{10}, \]
where $\alpha_{10}$ is a real constant. Now all the determining equations are satisfied.
We have five arbitrary constants; $c_1, c_4, c_7, c_8,$ and $c_{11}$. The integrals of motion corresponding to $c_7$ and $c_8$ are trivial. The non-trivial integrals of motion are $Y_s^{15}$, $Y_s^{16}$, and
\begin{equation}
     Y_s^{17}=(\vec{\sigma}_1+\vec{\sigma}_2, \vec{L})-\dfrac{\hbar}{r^2}(\vec{\sigma}_1, \vec{x})(\vec{\sigma}_2, \vec{x}). \label{IoM17}
\end{equation}

The second case is $c_{11}=4c_4$. Then all the determining equations are satisfied for  an arbitrary $V_0$. We have four arbitrary constants, but only two of them provide non-trivial integrals, which are given by $Y_s^{17}$ and 
\begin{equation}
    Y_s^{18}=4 Y_s^{16}+Y_s^{15}. \label{IoM18}
\end{equation}

\paragraph{B. $c_{11}=0$.} In this case all the determining equations obtained by the second-order terms are zero. Then we go on to analyze the determining equations obtained from the first- and zeroth-order terms. 
From \eqref{feq9} we have
\begin{equation}
    c_3\left( \hbar^2\left( \alpha_6+\dfrac{1}{2r^2} \right)-\hbar V_1+4V_2+2r^2 V_3 \right)=0. \label{f3eq9}
\end{equation}
Two cases arise from this equation.
\paragraph{a. $c_3\neq 0$.} Solving $\hbar^2\left( \alpha_6+\dfrac{1}{2r^2} \right)-\hbar V_1+4V_2+2r^2 V_3=0$ for $V_3$, we get
\[ V_3=-\dfrac{\hbar^2}{4r^4}-\dfrac{\alpha_6\hbar^2-\hbar V_1+4 V_2}{2r^2}. \]
We introduce these relations into equation \eqref{feq5}, and we obtain
\begin{align}
   &\dfrac{\hbar(\hbar(c_1+c_2)(1+2\alpha_6 r^2)-2(c_7-c_8-c_9)(5+6\alpha_6 r^2))}{2r} \nonumber\\
   &\qquad\quad-(2(-c_7+c_8+c_9)+\hbar(c_1+c_2))r V_1+4(c_1+c_2)r V_2=0. \label{f4eq5}
\end{align}
We have two options to analyze.
\paragraph{1. $c_1+c_2\neq0$.} We can solve \eqref{f4eq5} for $V_2$ to get
\begin{align}
    V_2=&\dfrac{-\hbar(\hbar(c_1+c_2)(1+2\alpha_6 r^2)-2(c_7-c_8-c_9)(5+6\alpha_6 r^2))}{8(c_1+c_2)r^2}\nonumber\\
    & +\dfrac{(2(-c_7+c_8+c_9)+\hbar(c_1+c_2))r V_1}{4(c_1+c_2)r}.
\end{align}
Next from \eqref{zeq1}
\begin{align}\label{zeq1n}
    c_4\left( 2(-c_7+c_8+c_9)(5\hbar+r^3 V_1')+3\hbar(c_1+c_2)(\hbar+r^3 V_1')-4(c_1+c_2)r^3 V_0' \right)=0.
\end{align}
We are left with two possibilities to consider. The first one is $c_4\neq 0$. Then solving for $V_0$ the differential equation obtained by equating the second factor on the left-hand side of \eqref{zeq1n}, since $c_1+c_2\neq 0$ we find that
\begin{equation}
    V_0=-\dfrac{\hbar(10(-c_7+c_8+c_9)+3\hbar(c_1+c_2))}{8(c_1+c_2)r^2}+\dfrac{(2(-c_7+c_8+c_9)+3\hbar(c_1+c_2))V_1}{4(c_1+c_2)}+\alpha_{11},
\end{equation}
where $c_{11}$ is a real constant. 
Now all the determining equations are satisfied for an arbitrary $V_1$. The integral of motion depends on $c_1, c_2, c_3, c_4, c_7, c_8$, and $c_9$. However the potentials $V_0, V_2$, and $V_3$ also depend on $\beta=(-c_7+c_8+c_9)/(c_1+c_2)$. We can choose $c_9=c_7-c_8+\beta(c_1+c_2)$, and so we have the following non-trivial integrals of motion for this case: $Y_s^9, Y_s^{10}, Y_s^{11},$ and $Y_s^{15}$.

The second possibility is $c_4=0$. Then all the determining equations are satisfied for arbitrary $V_0$ and $V_1$. Now we have the same integrals of motion but $Y_s^{15}$ as in the previous case.
\paragraph{2. $c_1+c_2=0$.} From \eqref{feq1}, we find that
\[ (c_7-c_8-c_9)(5\hbar +6\hbar \alpha_6 r^2-2r^2 V_1 )=0. \]
Notice here that we already have the assumption $5\hbar +6\hbar \alpha_6 r^2-2r^2 V_1\neq 0$.
Therefore $c_9=c_7-c_8$. Introducing these relations into \eqref{zeq1} and \eqref{zeq2}
\[ c_4(\hbar^2 -2r^3 V_0'+\hbar r^3 V_1'+2r^3 V_2')=0. \]
We address the following aspects. Firstly suppose that $\hbar^2 -2r^3 V_0'+\hbar r^3 V_1'+2r^3 V_2'=0$. Solving this equation for $V_2$ yields 
\[ V_2=\dfrac{\hbar^2}{4r^2}+V_0-\dfrac{\hbar}{2}V_1+\alpha_{12}, \]
where $\alpha_{12}$ is an integration constant. Now all the determining equations are satisfied for arbitrary $V_0$ and $V_1$. We have five arbitrary constants, only three of which give non-trivial integrals of motion. These are $Y_s^8$, $Y_s^9$, and $Y_s^{15}$.

Secondly suppose that $c_4=0$. In this case all the determining equations are satisfied for arbitrary $V_0, V_1,$ and $V_2$. We have two non-trivial integrals of motion, $Y_s^8$ and $Y_s^9$.

\paragraph{b. $c_3=0$.} Using \eqref{feq1}, we get
\[ (c_7-c_8-c_9)(5\hbar+6\hbar \alpha_6r^2-2r^2 V_1)+2(c_1+c_2)r^4 V_3=0.  \]
To move on the analysis, we need to consider two cases. 
\paragraph{1. $c_1+c_2\neq0$.} In this case solving the last equation for $V_3$
\[ V_3=\dfrac{\beta (5\hbar+6\hbar \alpha_6r^2-2r^2 V_1)}{2r^4}, \]
where $\beta=-(c_7-c_8-c_9)/(c_1+c_2)$. Now from \eqref{feq5} 
\begin{align*}
   & (c_1-c_2)\big(-2(c_7-c_8-c_9)(5\hbar+6\alpha_6\hbar r^2-2r^2 V_1)+\hbar(c_1+c_2)(\hbar-2r^2 V_1)\\ &\qquad\qquad+8(c_1+c_2)r^2 V_2\big)=0.
\end{align*}
This equation implies two circumstances. 
\paragraph{i. $c_1-c_2\neq0$.}
Now from the last equation, we immediately have
\[ V_2=-\dfrac{\hbar(\hbar+2\alpha_6\hbar r^2+2\beta(5+6\alpha_6 r^2))}{8r^2}+\dfrac{(2\beta + \hbar)}{4}V_1. \]
Then using \eqref{zeq1}, we get
\begin{align*}
    & c_4(\hbar(10(-c_7+c_8+c_9)+3\hbar(c_1+c_2))-4(c_1+c_2)r^3V_0' +(2(-c_7+c_8+c_9)\\ &\qquad +3\hbar(c_1+c_2))r^3 V_1')=0.
\end{align*}
Again we have two alternatives to analyze. Suppose that $c_4\neq 0$. Then from the last equation, we find that
\begin{equation*}
    V_0=-\dfrac{\hbar(10\beta+3\hbar)}{8r^2}+\dfrac{2\beta+3\hbar}{4}V_1+\alpha_{13},
\end{equation*}
where $\alpha_{13}$ is a real constant. So all the determining equations are satisfied for an arbitrary $V_1$. We have five arbitrary constants, but only three of them give non-trivial integrals of motion. These non-trivial integrals of motion are $Y_s^{10}$, $Y_s^{11}$, and $Y_s^{15}$.

Now assume that $c_4=0$. Then all the determining equations are satisfied for arbitrary $V_0$ and $V_1$. We have four arbitrary constants and only two non-trivial integrals of motion that are $Y_s^{10}$ and $Y_s^{11}$.
\paragraph{ii. $c_2=c_1$.} In this case, from equations \eqref{zeq1} and \eqref{zeq2} we obtain
\[ c_4( (-c_7+c_8+c_9)(5\hbar+r^3V_1') +2c_1r^3(V_0'-3V_2'))=0. \]
This equation leads us to two cases. Firstly assume that $(-c_7+c_8+c_9)(5\hbar+r^3V_1') +2c_1r^3(V_0'-3V_2')=0$. Solving this equation for $V_2$, we get
\[ V_2=-\dfrac{5\beta\hbar}{6r^2}+\dfrac{V_0}{3}+\dfrac{\beta V_1}{3}+\alpha_{14},\]
where $\alpha_{14}$ is a real constant. Note that $\beta=-(c_7-c_8-c_9)/(2c_1)$ and that $c_1\neq 0$ since in this case $c_1+c_2\neq 0$ and $c_2=c_1$. Now all the determining equations are satisfied for arbitrary $V_0$ and $V_1$. We have four arbitrary constant, but there are two non-trivial integrals of motion; $Y_s^{12}$ and $Y_s^{15}$.

Next suppose that $c_4=0$. In this case all the determining equations are satisfied for arbitrary $V_0, V_1,$ and $V_2$. We have only one non-trivial integral of motion, $Y_s^{12}$.
\paragraph{2. $c_1+c_2=0$.} Now from \eqref{feq1}, we obtain the following equation
\[ (c_7-c_8-c_9)(5\hbar + 6\alpha_6 r^2 -2r^2 V_1)=0. \]
Therefore we turn our attention to the following possibilities.
\paragraph{i. $5\hbar + 6\alpha_6 r^2 -2r^2 V_1=0$.}  Solving this equation for $V_1$
\[ V_1=\dfrac{5\hbar}{2r^2}+3\alpha_6\hbar. \]
Introducing these relations into equation \eqref{feq5}, we find that
\[ c_1 (\hbar^2 + \alpha_6 \hbar^2 r^2 - 2 r^2 V_2 - r^4 V_3)=0. \]
Then we need to consider two options. 
\paragraph{i$_1$. $\hbar^2 + \alpha_6 \hbar^2 r^2 - 2 r^2 V_2 - r^4 V_3=0$.} 
Then we solve this equation for $V_3$ to get
\[ V_3=\dfrac{\hbar^2+\alpha_6 \hbar^2 r^2-2r^2 V_2}{r^4}. \]
Now from \eqref{zeq1} we obtain
\[ c_4(2\hbar^2+r^3 (V_0'-V_2'))=0. \]
This equation results in the following possibilities. First, suppose that $2\hbar^2+r^3 (V_0'-V_2')=0$, which can be solved for $V_2$ to obtain 
\[ V_2=-\dfrac{\hbar^2}{r^2}+V_0+\alpha_{15},  \]
where $\alpha_{15}$ is a real constant. So all the determining equations are satisfied for arbitrary $V_0$. There are three non-trivial integrals of motion, $Y_s^6$, $Y_s^8$, and $Y_s^{15}$.

Now suppose that $c_4=0$. In this case all the determining equations are satisfied for arbitrary $V_0$ and $V_2$. There are two non-trivial integrals of motion, $Y_s^6$ and $Y_s^8$. 

\paragraph{i$_2$. $c_1=0$.} Introducing these relations into \eqref{zeq1}, we find that
\[ c_4(2r V_3-V_0'+3V_2'+r^2 V_3')=0.
\]
We now follow the two paths suggested by this equation. The first one is $2r V_3-V_0'+3V_2'+r^2 V_3'=0$. Solving this equation for $V_3$, we get 
\[ V_3=\dfrac{V_0-3V_2+\alpha_{16}}{r^2}, \]
where $\alpha_{16}$ is an integration constant. Now all the determining equations are satisfied for arbitrary $V_0$ and $V_2$. We have two non-trivial integrals of motion that correspond to the constants $c_4$ and $c_7$; $Y_s^6$ and $Y_s^{15}$.

The second case is $c_4=0$. Then all the determining equations are satisfied for arbitrary $V_0$, $V_2$, and $V_3$. There is only one non-trivial integral of motion, $Y_s^6$.

\paragraph{ii. $c_7-c_8-c_9=0$.} We directly have $c_9=c_7-c_8$. From \eqref{feq5}
\[ c_1(\hbar^2 +2 \alpha_6\hbar^2 r^2-2\hbar r^2 V_1+8r^2 V_2+4r^4V_3)=0.\]
We then proceed by investigating the following options. 

\paragraph{ii$_1$. $\hbar^2 +2 \alpha_6\hbar^2 r^2-2\hbar r^2 V_1+8r^2 V_2+4r^4V_3=0$.} Solving this equation for $V_3$
\[ V_3=-\dfrac{\hbar^2 +2 \alpha_6\hbar^2 r^2-2\hbar r^2 V_1+8r^2 V_2}{4r^4}. \]
Then using \eqref{zeq1} we get
\[ c_4 (\hbar^2-2r^2 V_0'+\hbar r^3 V_1'+2r^3 V_2')=0. \]
Firstly suppose that $\hbar^2-2r^2 V_0'+\hbar r^3 V_1'+2r^3 V_2'=0$. From this equation, we have
\[ V_2=\dfrac{\hbar^2}{4r^2}+V_0-\dfrac{\hbar}{2} V_1+\alpha_{17},\]
where $\alpha_{17}$ is a real constant. Therefore, all the determining equations are satisfied for arbitrary $V_0$ and $V_1$. We have four arbitrary constants; $c_1, c_4, c_7, c_8$ and two non-trivial integrals of motion; $Y_s^8$ and $Y_s^{15}$.

Secondly suppose that $c_4=0$. Then all the determining equations are satisfied for arbitrary $V_0, V_1,$ and $V_2$. In this case, we have only one non-trivial integral of motion, $Y_s^8$. 

\paragraph{ii$_2$. $c_1=0$.} In this case from \eqref{zeq1} we find that
\[ c_4(2r V_3- V_0'+3V_2'+r^2 V_3')=0. \]
Hence we have two cases again. 

Suppose that $2r V_3- V_0'+3V_2'+r^2 V_3'=0$. Solving this equation for $V_3$, we get
\[ V_3=\dfrac{V_0-3V_2+\alpha_{18}}{r^2}, \]
where $\alpha_{18}$ is an integration constant. Then all the determining equations are satisfied for arbitrary $V_0, V_1,$ and $V_2$. There is only one non-trivial integral of motion, $Y_s^{15}$.

Now suppose that $c_4=0$. Then all the determining equations are satisfied for arbitrary $V_0, V_1, V_2,$ and $V_3$, but no non-trivial integral of motion is found. 

\paragraph{II. $5\hbar+6\alpha_6 \hbar r^2-2r^2 V_1=0.$} This equation immediately yields
\[ V_1=\dfrac{5\hbar}{2r^2}+3\alpha_6 \hbar. \]
Then from \eqref{seq1} and \eqref{seq5}, we find that
\[ c_{11}=0,\qquad f_{15}=c_{12}, \]
where $c_{12}$ is a real constant. So all the determining equations coming from the second-order terms are satisfied. From \eqref{feq1} and \eqref{feq2},
\begin{equation}\label{fe12to}
     (c_1+c_2)V_3,\qquad (2c_{12}+c_3)V_3=0.
\end{equation}
Hence we examine the following situations: $V_3=0$ and $V_3\neq 0$.
\paragraph{A. $V_3=0$.} In this case, from \eqref{feq5}
\[ (c_1-c_2-\mathrm{i}c_3)(\hbar^2 + \alpha_6 \hbar^2 r^2-2r^2 V_2)=0. \]
This equation leaves us with the following alternatives.
\paragraph{a. $\hbar^2 + \alpha_6 \hbar^2 r^2-2r^2 V_2=0$.} Solving this equation for $V_2$, we obtain
\[ V_2=\dfrac{\hbar^2}{2r^2}+\dfrac{\alpha_6 \hbar^2}{2}. \]
Using \eqref{zeq1}, we find the following equation
\[ c_4(3\hbar^2 + r^3 V_0')=0. \]
Hence we consider two situations. The first one is $3\hbar^2 + r^3 V_0'=0$. This equation can be solved to get
\[ V_0=\dfrac{3\hbar^2}{2r^2}+\alpha_{19}, \]
where $\alpha_{19}$ is a real constant. Hence all the determining equations are satisfied. There are eight arbitrary constants. Two of these constants give trivial integrals of motion. The remaining constants give the following non-trivial integrals of motion: $Y_s^1$, $Y_s^2$, $Y_s^3$, $Y_s^5$, $Y_s^6$, and $Y_s^{15}$.

The second one is $c_4=0$. Then all the determining equations are satisfied for arbitrary $V_0$. In this case we have the same integrals of motion but $Y_s^{15}$ as in the previous case.
\paragraph{b. $c_2=c_1$, $c_3=0$.} From \eqref{zeq1}
\[ c_4(V_0'-3V_2')=0. \]
This equation yields the following cases. The first case is $V_0'-3V_2'=0$. From this equation we get
\[ V_2=\dfrac{V_0}{3}+\alpha_{20}, \]
where $\alpha_{20}$ is a real constant. Now all the determining equations are satisfied for arbitrary $V_0$. There are six arbitrary constants, but only four non-trivial integrals of motion exist: $Y_s^5, Y_s^6, Y_s^7$, and $Y_s^{15}$.

The second case is $c_4=0$. Then all the determining equations are satisfied for arbitrary $V_0$ and $V_2$. We have the same integrals of motion except $Y_s^{15}$ as in the previous case.

\paragraph{B. $V_3\neq 0$.} Then it follows from \eqref{fe12to} that $c_2=-c_1$ and $c_3=-2c_{12}$. Introducing these relations into \eqref{feq5} yields
\begin{equation}\label{feq5den}
    (\mathrm{i}c_1+c_{12})(\hbar^2+\alpha_6\hbar^2 r^2-2r^2 V_2-r^4V_3)=0.
\end{equation}
Therefore we need to consider two instances.
\paragraph{a. $\hbar^2+\alpha_6\hbar^2 r^2-2r^2 V_2-r^4V_3=0$.} Solving this equation for $V_3$, we get
\[ V_3=\dfrac{\hbar^2+\alpha_6 \hbar^2 r^2-2r^2 V_2}{r^4}. \]
Now from \eqref{zeq1}
\[ c_4(2\hbar^2+r^3(V_0'-V_2)')=0. \]
This means that we have two cases to analyze. The first case is $2\hbar^2+r^3(V_0'-V_2)'=0$, which directly yields
\[ V_2=-\dfrac{\hbar^2}{r^2}+V_0+\alpha_{21}, \]
where $\alpha_{21}$ is a real constant. Hence all the determining equations are satisfied for arbitrary $V_0$. The non-trivial integrals of motion in this case are $Y_s^6, Y_s^8,Y_s^9$, and $Y_s^{15}$. 

The second case is $c_4=0$. Now all the determining equations are satisfied for arbitrary $V_0$ and $V_2$. The non-trivial integrals of motion are $Y_s^6, Y_s^8$, and $Y_s^9$.
\paragraph{b. $\hbar^2+\alpha_6\hbar^2 r^2-2r^2 V_2-r^4V_3\neq0$.} In this case from \eqref{feq5den} we have $c_1=0$ and $c_{12}=0$. Then using \eqref{zeq1} we get
\[ c_4(2rV_3-V_0'+3V_2'+r^2 V_3')=0,  \]
which yields the following two possibilities. The first possibility is $2rV_3-V_0'+3V_2'+r^2 V_3'=0$. From this equation 
\[ V_3=\dfrac{V_0-3V_2+\alpha_{22}}{r^2}, \]
where $\alpha_{22}$ is a real constant. Then all the determining equations are satisfied for arbitrary $V_0$ and $V_2$. In this case we have two non-trivial integrals of motion, $Y_s^6$ and $Y_s^{15}$. 

The second possibility is $c_4=0$. Then all the determining equations are satisfied for arbitrary $V_0$, $V_2$, and $V_3$. In this case we have only one non-trivial integral of motion that is $Y_s^6$.
\paragraph{S2. $c_4=0$.} We go back to the determining equations coming from the second-order terms. From \eqref{seq14}, we find that
\[ \hbar f_{10}+(c_3+2f_5)(V_1-3\hbar V_5)=0. \]
This equation gives us
\[ f_{10}=-\dfrac{(c_3+2f_5)(V_1-3\hbar V_5)}{\hbar}. \]
Introducing these relations into \eqref{seq16} yields
\[ (c_3+2f_5)(\hbar-r^2 V_1+3\hbar r^2 V_5)(1+r^3V_5')=0. \]
This equation suggests three cases, but the case $1+r^3V_5'=0$ has already been discussed previously. Therefore we shall consider the remaining two cases.
\paragraph{I. $\hbar-r^2 V_1+3\hbar r^2 V_5=0$.} We solve this equation for $V_5$ and then obtain
\[ V_5=-\dfrac{1}{3r^2}+\dfrac{V_1}{3\hbar}. \]
Substituting this relation into the remaining determining equations originating from the second-order terms, we get $f_5'=0$, that is, $f_5=c_{13}$, where $c_{13}$ is a real constant. Now all of these equations are satisfied, and so we move on to analyze the determining equations coming from the first- and the zeroth-order terms. Using \eqref{feq1} and \eqref{feq2}, we see that
\begin{equation}\label{fe12to2}
     (c_1+c_2)V_3,\qquad (2c_{12}+c_3)V_3=0.
\end{equation}
Therefore there are two distinct paths we can follow: $V_3=0$ and $V_3\neq 0$.
\paragraph{A. $V_3=0$.} From \eqref{feq5}
\[ (c_1-c_2-\mathrm{i}c_3)(\hbar^2+2\hbar r^2 V_1-12r^2 V_2)=0. \]
Hence we base our further analysis on the following two options.
\paragraph{a. $\hbar^2+2\hbar r^2 V_1-12r^2 V_2=0$.} Solving this equation for $V_2$, we get
\[ V_2=\dfrac{\hbar^2}{12r^2}+\dfrac{\hbar}{6}V_1. \]
Then all the determining equations are satisfied for arbitrary potentials $V_0$ and $V_1$. In this case the non-trivial integrals of motion are $Y_s^1, Y_s^2, Y_s^3, Y_s^5,$ and $Y_s^6$. 
\paragraph{b. $\hbar^2+2\hbar r^2 V_1-12r^2 V_2\neq0$.} In this case we have $c_3=0$ and $c_2=c_1$. Then all the determining equations are satisfied for arbitrary $V_0, V_1,$ and $V_2$. The non-trivial integrals of motion are $Y_s^5,Y_s^6,$ and $Y_s^7$. 
\paragraph{B. $V_3\neq0$.} In this case from \eqref{fe12to2} we have $c_2=-c_1$ and $c_{12}=-c_3/2$. Then from \eqref{feq5} we get
\[ (2\mathrm{i}c_1+c_3)(-\hbar^2-2\hbar r^2V_1+12r^2V_2+6r^4 V_3)=0, \]
which yields the following two cases.
\paragraph{a. $-\hbar^2-2\hbar r^2V_1+12r^2V_2+6r^4 V_3=0$.} This equation can be solved for $V_3$ to get
\[ V_3=\dfrac{\hbar^2+2\hbar r^2 V_1-12 r^2 V_2}{6r^4}. \]
Then all the determining equations are satisfied for arbitrary $V_0$, $V_1$, and $V_2$. We have three non-trivial integrals of motion that are $Y_s^6, Y_s^8,$ and $Y_s^9$.
\paragraph{b. $-\hbar^2-2\hbar r^2V_1+12r^2V_2+6r^4 V_3\neq0$.} In this case we have $c_1=0$ and $c_3=0$. Then all the determining equations are satisfied for arbitrary potentials $V_0, V_1, V_2,$ and $V_3$. We have only one non-trivial integral of motion, $Y_s^6$. 
\paragraph{II. $f_5=-c_3/2$.} In this case using \eqref{feq1} and \eqref{feq5} we obtain 
\begin{equation}\label{fe15to3}
    (\mathrm{i}(c_1-c_2)+c_3)(-\hbar V_1+4V_2+2r^2 V_3+\hbar^2 V_5).
\end{equation}
We carry on our analysis with the following two possibilities.
\paragraph{A. $-\hbar V_1+4V_2+2r^2 V_3+\hbar^2 V_5=0$.} We solve this equation for $V_5$ and find that
\[ V_5=\dfrac{\hbar V_1-4V_2-2r^2 V_3}{\hbar^2}. \]
Then from \eqref{feq1} we get
\[ (c_1+c_2)r^2 V_3+\dfrac{(c_7-c_8-c_9)(\hbar^2+2\hbar r^2V_1-12r^2 V_2-6r^4 V_3)}{\hbar r^2}. \]
Hence we shall consider the following options.
\paragraph{a. $6(-c_7+c_8+c_9)+\hbar(c_1+c_2)\neq 0$.} Then solving the last equation for $V_3$, we find that
\[ V_3=\dfrac{(-c_7+c_8+c_9)(\hbar^2+2\hbar r^2 V_1-12r^2 V_2)}{(6(-c_7+c_8+c_9)+\hbar(c_1+c_2))r^4}. \]
Then all the determining equations are satisfied for arbitrary $V_0, V_1,$ and $V_2$. Set $\beta_1=c_7-c_8-c_9$ and $\beta_2=c_1+c_2$, which yields $c_9=c_7-c_8-\beta_1$ and $c_2=\beta_2-c_1$. Then we have four arbitrary constants, but two non-trivial integral of motion $Y_s^8$ and $Y_s^9$. 
\paragraph{b. $6(-c_7+c_8+c_9)+\hbar(c_1+c_2)=0$.} Now we have $c_9=c_7-c_8-\hbar(c_1+c_2)/6$.
Then from \eqref{feq1} 
\[ (c_1+c_2)(\hbar^2+2\hbar r^2 V_1-12r^2 V_2)=0. \]
The analysis progresses by considering the following cases. The first case is $\hbar^2+2\hbar r^2 V_1-12r^2 V_2=0$. Solving this equation for $V_2$
\[ V_2=\dfrac{\hbar^2}{12r^2}+\dfrac{\hbar}{6}V_1. \]
Hence all the determining equations are satisfied for arbitrary $V_0$ and $V_1$. We have three non-trivial integrals of motion, $Y_s^9$,
\begin{align}
       &Y_s^{19}=(\vec{\sigma}_2, \vec{L})-\dfrac{\hbar}{6r^2}(\vec{\sigma}_1,\vec{x})(\vec{\sigma}_2,\vec{x}), \label{IoM19} \\
     &Y_s^{20}=(\vec{\sigma}_1, \vec{L})-\dfrac{\hbar}{6r^2}(\vec{\sigma}_1,\vec{x})(\vec{\sigma}_2,\vec{x}). \label{IoM20}
\end{align}

The second case is $c_2=-c_1$. Then all the determining equations are satisfied for arbitrary potentials $V_0, V_1, V_2,$ and $V_3$. In this case we have two non-trivial integrals of motion that are $Y_s^8$ and $Y_s^9$.
\paragraph{B. $-\hbar V_1+4V_2+2r^2 V_3+\hbar^2 V_5\neq0$.} Now from \eqref{fe15to3} we find that $c_2=c_1$ and $c_3=0$. Then using \eqref{feq1} we obtain
\begin{equation}\label{feq1new}
    (-c_7 + c_8 + c_9)(-\hbar+ r^2 V_1-3r^2 V_5)+ 
 2 c_1 r^4 V_3=0. 
\end{equation}
The following states arise from this equation.
\paragraph{a. $-c_7 + c_8 + c_9\neq 0$.} Then from the last equation we get
\[ V_5=\dfrac{(c_7-c_8-c_9)(-\hbar+r^2 V_1)-2c_1 r^4 V_3}{3(c_7-c_8-c_9)\hbar r^2}. \]
Then all the determining equations are satisfied for arbitrary $V_0, V_1, V_2,$ and $V_3$. Set $\beta=c_7-c_8-c_9$. Then we have $c_9=c_7-c_8-\beta$. In this case there are two arbitrary constants, $c_7$ and $c_8$, but the corresponding integrals of motion are trivial.
\paragraph{b. $c_9=c_7-c_8$.} From \eqref{feq1new} we get $c_1V_3=0$. This leads us to the following cases. The first case is $V_3=0$. In this case all the determining equations are satisfied for arbitrary $V_0,V_1, V_2,$ and $V_5$. The only non-trivial integral of motion is $Y_s^7$.

The second case is $c_1=0$. Then for arbitrary potentials $V_0, V_1, V_2, V_3,$ and $V_5$, all the determining equations are satisfied. But there is no non-trivial integral of motion in this case.

\subsection*{Case 2. $V_4\neq 0$}
From \eqref{deteq3rdorder}, we conclude that
\begin{equation}
    f_5=f_6=f_8=f_9=f_{10}=0.
\end{equation}
So all the determining equations coming from the third-order terms are satisfied. Now we consider the determining equations obtained by setting the coefficients of the second-order terms to zero. These terms give us the following equations:
\begin{align}
    &f_2+f_7=0, \label{c2seq1}\\
    &f_4=0, \label{c2seq2} \\
    &(-1+2V_4)f_7'+r^2f_7V_5'=0, \label{c2seq3}\\
    & f_7(1-2V_4+rV_4'+r^3 V_5')=0. \label{c2seq4}
\end{align}
Based on \eqref{c2seq4}, we proceed by analyzing the following subcases.
\subsubsection*{Subcase 1. $f_7\neq 0$}
In this case, from \eqref{c2seq4} we have $1-2V_4+rV_4'+r^3 V_5'=0$.
Solving this equation for $V_5$ yields
\begin{equation}
    V_5=\dfrac{1}{2r^2}-\dfrac{V_4}{r^2}+\alpha_{23},
\end{equation}
where $\alpha_{23}$ is a real constant. Introducing this relation into \eqref{c2seq3}, we find that 
\[ r(-1+2V_4)f_7'+f_7(-1+2V_4-rV_4')=0. \]
Since $f_7\neq0$, this equation leads to
\begin{equation}
    V_4=\dfrac{1}{2}+\alpha_{24}r^2f_7^2,
\end{equation}
where $\alpha_{24}$ is a real constant. Next we consider the determining equations coming from the coefficients of the first-order and the zeroth-order terms. We have the following equations:
\begin{align}
    & f_1'+f_3'=0, \label{c2feq1}\\
    & \alpha_{24}f_7^2 (f_1'+\mathrm{i}\hbar(2f_7'+rf_7''))=0, \label{c2feq2}\\
    & f_7\big( 4r V_3-2V_0'+6V_2'+2r^2V_3'+\alpha_{24}\hbar f_7(-4\mathrm{i}f_7'+8\hbar f_7'+r(-2\mathrm{i}f_1''+7\hbar f_7''+\hbar r f_7''' )) \big)=0. \label{c2feq3}
\end{align}
From \eqref{c2feq1}, we get $f_3=-f_1+c_{13}$, where $c_{13}$ is a real constant. Since $f_7\neq0$, by \eqref{c2feq2}, we see that $\alpha_{24}=0$ or $f_1'+\mathrm{i}\hbar(2f_7'+rf_7'')=0$. Our analysis continues based on the following two options.
\paragraph{S1. $f_1'+\mathrm{i}\hbar(2f_7'+rf_7'')=0$.} In this case, we have $f_1=c_{14}$ and $f_7=-\dfrac{c_{15}}{r}+c_{16}$, where $c_{14}$, $c_{15}$, and $c_{16}$ are real constants and $(c_{15},c_{16})\neq (0,0)$. Substituting these relations in \eqref{c2feq3}, we find that
\[ (c_{15}-c_{16}r)(2r V_3-V_0'+3V_2'+r^2V_3')=0. \]
Since $f_7\neq 0$, i.e., $(c_{15},c_{16})\neq (0,0)$, we obtain $2r V_3-V_0'+3V_2'+r^2V_3'=0$. We solve this equation for $V_3$ to get
\[ V_3=\dfrac{\alpha_{25}+V_0-3V_2}{r^2}, \]
where $\alpha_{25}$ is a real constant. Then all the determining equations are satisfied. We have two arbitrary constants, $c_{13}$ and $c_{14}$, which are not appearing in the Hamiltonian. However, the scalar integrals of motion corresponding to these constants are trivial ones.
\paragraph{S2. $\alpha_{24}=0$.} In this case we have $V_4=\dfrac{1}{2}$ and $V_5=\alpha_{23}$. Then all the determining equations coming from the first-order terms are satisfied. We have only one determining equation obtained from the coefficients of the zeroth-order terms, which is
\[ f_7(2rV_3-V_0'+3V_2'+r^2V_3')=0. \]
Since $f_7\neq 0$, we have $2rV_3-V_0'+3V_2'+r^2V_3'=0$ that again gives us
\[ V_3=\dfrac{\alpha_{26}+V_0-3V_2}{r^2}, \]
where $\alpha_{26}$ is a real constant. Now all the determining equations are satisfied. Notice that we have one arbitrary constant $c_{13}$ and the real functions $f_1$ and $f_7$ are arbitrary. The integral of motion corresponding to $c_{13}$ is the trivial one, $(\vec{\sigma}_1, \vec{\sigma}_2)$. Since $f_1$ and $f_7$ are arbitrary, we also have an infinite family of integrals of motion given by
\begin{equation}\label{IoM21}
    Y_s^{21}=\big( 1-(\vec{\sigma}_1, \vec{\sigma}_2) \big)\left( f_1+\dfrac{\mathrm{i}\hbar}{2}(r f_7'+3f_7)-f_7(\vec{x},\vec{p})  \right).
\end{equation}
\subsubsection*{Subcase 2. $f_7= 0$} Setting $f_7=0$ in \eqref{c2seq1} yields $f_2=0$. Then all the determining equations associated with the second-order terms are satisfied. Next we find the following determining equations associated with the first-order and the zeroth-order terms:
\begin{align}
    & f_1'+f_3'=0, \\
    & f_1'+(-1+4V_4)f_3'=0, \\
    & (-1+2V_4)f_3'=0. 
\end{align}
Fron the first equation we directly have $f_3=-f_1+c_{17}$, where $c_{17}$ is a real constant. Then we have $(-1+2V_4)f_1'=0$, which gives rise to the following two situations.
\paragraph{S1. $V_4=\dfrac{1}{2}$.} 
In this case all the determining equations are satisfied. We have one arbitrary constant $c_{18}$ and the corresponding scalar integral of motion is $(\vec{\sigma}_1, \vec{\sigma}_2)$. All the potentials in the Hamiltonian except $V_4$ are arbitrary. We also see that $f_1$ is an arbitrary function of $r$. This means that we again have an infinite family of integrals of motion given by
\begin{equation}\label{IoM22}
    Y_s^{22}=f_1(1-(\vec{\sigma}_1,\vec{\sigma}_2)).
\end{equation}
Note that even though $1-(\vec{\sigma}_1,\vec{\sigma}_2)$ is a trivial scalar integral of motion, unless $f_1\equiv 1$, the integral of motion given by \eqref{IoM22} will commute with the Hamiltonian only when $V_4=1/2$.
\paragraph{S2. $f_1'=0$.} We have $f_1=c_{19}$, where $c_{19}$ is a real constant. Then all the determining equations are satisfied. We see that all the potentials in the Hamiltonian are arbitrary. Therefore we have only two trivial integrals of motion for this case, which are $(\vec{\sigma}_1,\vec{\sigma_2})$ and $1-(\vec{\sigma}_1,\vec{\sigma_2})$. 

\section{Conclusions}
The systematic investigation and classification of superintegrable systems involving spin interaction were initiated in 2006 \cite{Winternitz.c}, and the classification was completed for such systems admitting up to second-order integrals of motion \cite{DWY, YTW}. However, in these papers, the problem is restricted to the interaction of two particles, only one of which has spin. Now, in this paper, we have generalized this problem and initiated a research program to systematically investigate the superintegrability of systems in the interaction of two particles both of which have spin $1/2$. Such systems can be interpreted as \textit{e.g.} a nucleon-nucleon interaction, which means that they have significant physical relevance and applications. We have first identified the form of the Hamiltonian required to explore these systems. Okubo and Marshak \cite{OM} proposed the most general potential for two nucleon interaction based on some physical restrictions, including charge independence, translational invariance, Galilean invariance, permutation symmetry, rotational invariance, space reflection invariance, time reversal invariance, and Hermiticity. We have clearly explained these restrictions on the potential and obtained the terms that the Hamiltonian must contain. Then the most general Hamiltonian is
\begin{align*}
    H=&-\dfrac{\hbar^2}{2}\Delta+V_0(r)+V_1(r)\dfrac{1}{2}(\vec{\sigma}_1+\vec{\sigma}_2,\vec{L})+V_2(r)(\vec{\sigma}_1,\vec{\sigma}_2)+V_3(r)(\vec{x},\vec{\sigma}_1)(\vec{x},\vec{\sigma}_2)\\
	&+V_4(r)(\vec{\sigma}_1,\vec{p})(\vec{\sigma}_2,\vec{p})+V_5(r)\dfrac{1}{2}\big((\vec{\sigma}_1,\vec{L})(\vec{\sigma}_2,\vec{L})+(\vec{\sigma}_2,\vec{L})(\vec{\sigma}_1,\vec{L})\big).
\end{align*}
Before delving into the analysis, we have identified the potentials that we should exclude from the analysis. More specifically, we have determined what specific choice of the Hamiltonian can be directly derived by applying a gauge transformation to the scalar Hamiltonian. And we have found that the following set of potentials is gauge induced
\begin{equation*}
V_0=V_0(r),\quad  V_1=\dfrac{2\hbar}{r^2},\quad V_2=\dfrac{\hbar^2}{r^2},\quad V_3=-\dfrac{\hbar^2}{r^4},\quad V_4=0,\quad V_5=0.
\end{equation*}
We have restricted our analysis to first-order scalars. That is, we have investigated superintegrable systems in the interaction of two particles with spin admitting first-order scalar integrals of motion. Then we have realized that interestingly there are several non-trivial integrals of motion and superintegrable systems unlike the case with systems involving two particles only one of which has spin \cite{wy3}. The main results of this paper are encapsulated in the following theorem.
\begin{theorem}
Assume that all $c_i$'s, $\alpha_j$'s, and $\beta_k$'s given below are real constants and $\epsilon^2=1$. The only spherically symmetric superintegrable systems with spin admitting first-order scalar integrals of motion are the following:
    \begin{enumerate}
        \item $V_0=\alpha_2$, $V_2=\alpha_1\hbar^2/2$, $V_3=0$, $V_4=0$,
        \[ V_1=-\dfrac{\hbar}{2r^2}+3\alpha_1\hbar,\quad V_5=-\dfrac{1}{2r^2}+\alpha_1. \]
        There are six non-trivial scalar integrals of motion given by \eqref{IoM1}, \eqref{IoM2}, \eqref{IoM3}, \eqref{IoM4}, \eqref{IoM5}, and \eqref{IoM6}.
        \item $V_0=V_0(r)$, $V_2=-V_0+\alpha_3$, $V_3=0$, $V_4=0$,
        \[ V_1=-\dfrac{\hbar}{2r^2}+3\alpha_1\hbar,\quad V_5=-\dfrac{1}{2r^2}+\alpha_1. \]
        There are four non-trivial scalar integrals of motion given by \eqref{IoM4}, \eqref{IoM5}, \eqref{IoM6}, and \eqref{IoM7}.
        \item $V_0=V_0(r)$, $V_4=0$,
        \begin{align*}
        & V_3=\dfrac{\alpha_1\hbar^2-2\alpha_4}{r^2}+\dfrac{2}{3r^2}V_0,\quad V_2=-\dfrac{V_0}{3}+\alpha_4,\\
           & V_1=-\dfrac{\hbar}{2r^2}+3\alpha_1\hbar,\quad V_5=-\dfrac{1}{2r^2}+\alpha_1.
        \end{align*}
         There are four non-trivial scalar integrals of motion given by \eqref{IoM4}, \eqref{IoM6}, \eqref{IoM8}, and \eqref{IoM9}.
         \item $V_0=V_0(r), V_2=V_2(r)$,
        \[ V_4=0,\quad V_3=\dfrac{V_0+V_2+\alpha_5}{r^2},\quad V_1=-\dfrac{\hbar}{2r^2}+3\alpha_1\hbar,\quad V_5=-\dfrac{1}{2r^2}+\alpha_1. \]
        There are two non-trivial scalar integrals of motion given by \eqref{IoM4} and \eqref{IoM6}.
        \item $V_0=V_0(r), V_1=V_1(r)$, $V_4=0$,
        \begin{align*}
            &V_5=-\dfrac{1}{2r^2}+\alpha_1,\quad V_3=\dfrac{\beta(-\hbar+6\alpha_1\hbar r^2-2r^2 V_1)}{2r^4}, \\
            & V_2=\dfrac{\hbar(2\beta+\hbar)}{8r^2}+\dfrac{2\beta+\hbar}{4}V_1-\dfrac{\alpha_1\hbar}{4}(6\beta+\hbar).
        \end{align*}
        There are three non-trivial scalar integrals of motion given by \eqref{IoM9}, \eqref{IoM10}, and \eqref{IoM11}.
        \item $V_0=V_0(r), V_1=V_1(r), V_2=V_2(r)$, $V_4=0$,
        \[ V_5=-\dfrac{1}{2r^2}+\alpha_1,\quad V_3=\dfrac{\hbar^2-2\alpha_1\hbar^2 r^2+2\hbar r^2 V_1-8r^2 V_2}{4r^4}.  \]
        There are two non-trivial scalar integrals of motion given by \eqref{IoM8} and \eqref{IoM9}.
        \item $V_0=V_0(r), V_1=V_1(r), V_2=V_2(r)$, $V_4=0$,
        \[ V_5=-\dfrac{1}{2r^2}+\alpha_1,\quad V_3=\dfrac{\beta(-\hbar+6\alpha_1\hbar r^2-2r^2 V_1)}{2r^4}.  \]
        There is only one non-trivial scalar integral of motion given by \eqref{IoM12}.
        \item $V_5=1/(2r^2)+\alpha_6$, $V_4=0$,
        \begin{align*}
            & V_1=\dfrac{5\hbar}{2r^2}+3\alpha_6\hbar+\dfrac{\epsilon \hbar}{r^2\sqrt{2+\alpha_7\hbar^2 r^2}},\quad V_3=-\dfrac{\epsilon \hbar^2}{2r^4\sqrt{2+\alpha_7\hbar^2 r^2}}, \\
            &V_2=\dfrac{\hbar^2}{2r^2}+\dfrac{\epsilon\hbar^2}{2r^2 \sqrt{2+\alpha_7\hbar^2 r^2}}+\dfrac{\hbar^2 \alpha_6}{2},\quad V_0=\dfrac{3\hbar^2}{2r^2}+\dfrac{\epsilon\hbar^2}{r^2\sqrt{2+\alpha_7\hbar^2 r^2}}+\alpha_8.
        \end{align*}
        There are five non-trivial scalar integrals of motion given by \eqref{IoM9}, \eqref{IoM13}, \eqref{IoM14}, \eqref{IoM15}, and \eqref{IoM16}.
        \item $V_5=1/(2r^2)+\alpha_6$, $V_4=0$,
        \begin{align*}
            & V_1=\dfrac{5\hbar}{2r^2}+3\alpha_6\hbar+\dfrac{\epsilon \hbar}{r^2\sqrt{2+\alpha_7\hbar^2 r^2}},\quad V_3=-\dfrac{\epsilon \hbar^2}{2r^4\sqrt{2+\alpha_7\hbar^2 r^2}}, \\
            &V_2=\dfrac{\hbar^2}{2r^2}+\dfrac{\epsilon\hbar^2}{2r^2 \sqrt{2+\alpha_7\hbar^2 r^2}}+\alpha_9-\alpha_{10},\quad V_0=\dfrac{3\hbar^2}{2r^2}+\dfrac{\epsilon\hbar^2}{r^2\sqrt{2+\alpha_7\hbar^2 r^2}}+\alpha_{10}.
        \end{align*}
         There are three non-trivial scalar integrals of motion given by \eqref{IoM15}, \eqref{IoM16}, and \eqref{IoM17}.
         \item $V_0=V_0(r)$, $V_4=0$, $V_5=1/(2r^2)+\alpha_6$,
        \begin{align*}
            & V_1=\dfrac{5\hbar}{2r^2}+3\alpha_6\hbar+\dfrac{\epsilon \hbar}{r^2\sqrt{2+\alpha_7\hbar^2 r^2}},\quad V_3=-\dfrac{\epsilon \hbar^2}{2r^4\sqrt{2+\alpha_7\hbar^2 r^2}}, \\
            &V_2=\dfrac{2\hbar^2}{r^2}+\dfrac{3\epsilon\hbar^2}{2r^2 \sqrt{2+\alpha_7\hbar^2 r^2}}-V_0+\alpha_9.
        \end{align*}
        There are two non-trivial scalar integrals of motion given by \eqref{IoM17} and \eqref{IoM18}.
        \item $V_1=V_1(r)$, $V_4=0$,
        \begin{align*}
            & V_5=\dfrac{1}{2r^2}+\alpha_6,\quad V_3=\dfrac{\beta(5\hbar+6\alpha_6\hbar r^2-2r^2 V_1)}{2r^4}, \\
            & V_2=-\dfrac{\hbar(10\beta+\hbar)}{8r^2}+\dfrac{2\beta+\hbar}{4}V_1-\dfrac{\alpha_6\hbar}{4}(6\beta+\hbar),\quad V_0=-\dfrac{\hbar(10\beta+3\hbar)}{8r^2}+\dfrac{2\beta+3\hbar}{4}V_1+\alpha_{11}.
        \end{align*}
        There are four non-trivial scalar integrals of motion given by \eqref{IoM9}, \eqref{IoM10}, \eqref{IoM11}, and \eqref{IoM15}.
        \item $V_0=V_0(r), V_1=V_1(r)$, $V_4=0$,
        \begin{align*}
            & V_5=\dfrac{1}{2r^2}+\alpha_6,\quad V_3=\dfrac{\beta(5\hbar+6\alpha_6\hbar r^2-2r^2 V_1)}{2r^4}, \\
            & V_2=-\dfrac{\hbar(10\beta+\hbar)}{8r^2}+\dfrac{2\beta+\hbar}{4}V_1-\dfrac{\alpha_6\hbar}{4}(6\beta+\hbar).
        \end{align*}
        There are three non-trivial scalar integrals of motion given by \eqref{IoM9}, \eqref{IoM10}, and \eqref{IoM11}.
        \item $V_0=V_0(r), V_1=V_1(r)$, $V_4=0$,
        \begin{align*}
            & V_5=\dfrac{1}{2r^2}+\alpha_6,\quad V_3=-\dfrac{3\hbar^2+8\alpha_{12}r^2+2\alpha_6\hbar^2r^2+8r^2 V_0-6\hbar r^2 V_1}{4r^4}, \\
            & V_2=\dfrac{\hbar^2}{4r^2}+V_0-\dfrac{\hbar}{2}V_1+\alpha_{12}.
        \end{align*}
        There are three non-trivial scalar integrals of motion given by \eqref{IoM8}, \eqref{IoM9}, and \eqref{IoM15}.
        \item $V_0=V_0(r), V_1=V_1(r), V_2=V_2(r)$, $V_4=0$,
        \begin{align*}
            & V_5=\dfrac{1}{2r^2}+\alpha_6,\quad V_3=-\dfrac{3\hbar^2+8\alpha_{12}r^2+2\alpha_6\hbar^2r^2+8r^2 V_0-6\hbar r^2 V_1}{4r^4}.
        \end{align*}
        There are two non-trivial scalar integrals of motion given by \eqref{IoM8} and \eqref{IoM9}.
        \item $V_0=V_0(r), V_1=V_1(r)$, $V_4=0$,
        \begin{align*}
            & V_5=\dfrac{1}{2r^2}+\alpha_6,\quad V_3=\dfrac{\beta(5\hbar+6\alpha_6\hbar r^2-2r^2 V_1)}{2r^4}, \\
            & V_2=-\dfrac{5\beta \hbar}{6r^2}+\dfrac{V_0}{3}-\dfrac{\beta V_1}{3}+\alpha_{14}.
        \end{align*}
        There are two non-trivial scalar integrals of motion given by  \eqref{IoM12} and \eqref{IoM15}.
        \item $V_0=V_0(r), V_1=V_1(r), V_2=V_2(r)$, $V_4=0$,
        \begin{align*}
             V_5=\dfrac{1}{2r^2}+\alpha_6,\quad V_3=\dfrac{\beta(5\hbar+6\alpha_6\hbar r^2-2r^2 V_1)}{2r^4}.
        \end{align*}
        There is only one non-trivial scalar integral of motion given by  \eqref{IoM15}.
      \item   $V_0=V_0(r)$, $V_2=V_2(r)$,
        \begin{align*}
             &V_4=0,\quad V_5=\dfrac{1}{2r^2}+\alpha_6,\\
             &V_1=\dfrac{5\hbar}{2r^2}+3\alpha_6\hbar,\quad  V_3=\dfrac{V_0-3V_2+\alpha_{16}}{r^2}.
        \end{align*}
        There are two non-trivial scalar integrals of motion given by \eqref{IoM6} and \eqref{IoM15}.
        \item $V_0=V_0(r), V_1=V_1(r), V_2=V_2(r)$,
        \begin{align*}
            V_4=0,\quad V_5=\dfrac{1}{2r^2}+\alpha_6,\quad V_3=\dfrac{V_0-3V_2+\alpha_{18}}{r^2}.
        \end{align*}
        There is only one non-trivial scalar integral of motion given by  \eqref{IoM15}.
        \item $V_3=0$, $V_4=0$,
        \[ V_5=\dfrac{1}{2r^2}+\alpha_6,\quad V_1=\dfrac{5\hbar}{2r^2}+3\alpha_6\hbar,\quad V_2=\dfrac{\hbar^2}{2r^2}+\dfrac{\alpha_6\hbar^2}{2},\quad V_0=\dfrac{3\hbar^2}{2r^2}+\alpha_{19}.  \]
        There are six non-trivial scalar integrals of motion given by \eqref{IoM1}, \eqref{IoM2}, \eqref{IoM3}, \eqref{IoM5}, \eqref{IoM6}, and \eqref{IoM15}.
        \item $V_3=0$, $V_4=0$,
        \[ V_5=\dfrac{1}{2r^2}+\alpha_6,\quad V_1=\dfrac{5\hbar}{2r^2}+3\alpha_6\hbar,\quad V_2=\dfrac{V_0}{3}+\alpha_{20}.  \]
        There are four non-trivial scalar integrals of motion given by \eqref{IoM5}, \eqref{IoM6}, \eqref{IoM7}, and \eqref{IoM15}.
        \item $V_0=V_0(r)$, $V_4=0$,
        \begin{align*}
            &V_5=\dfrac{1}{2r^2}+\alpha_6,\quad V_1=\dfrac{5\hbar}{2r^2}+3\alpha_6\hbar, \\
            & V_3=\dfrac{-2\alpha_{20}r^2+3\hbar^2+\alpha_6\hbar^2 r^2-2r^2 V_0}{r^4},\quad V_2=-\dfrac{\hbar^2}{r^2}+V_0+\alpha_{21}.
        \end{align*}
        There are four non-trivial scalar integrals of motion given by \eqref{IoM6}, \eqref{IoM8}, \eqref{IoM9}, and \eqref{IoM15}.
        \item $V_0=V_0(r), V_1=V_1(r)$,
        \[ V_4=0,\quad V_3=0,\quad V_5=-\dfrac{1}{3r^2}+\dfrac{V_1}{3\hbar},\quad V_2=\dfrac{\hbar(\hbar+2r^2 V_1)}{12r^2}. \]
        There are five non-trivial scalar integrals of motion given by \eqref{IoM1}, \eqref{IoM2}, \eqref{IoM3}, \eqref{IoM5}, and \eqref{IoM6}.
        \item $V_0=V_0(r), V_1=V_1(r), V_2=V_2(r)$,
        \[ V_4=0,\quad V_3=0,\quad V_5=-\dfrac{1}{3r^2}+\dfrac{V_1}{3\hbar}. \]
        There are three non-trivial scalar integrals of motion given by \eqref{IoM5}, \eqref{IoM6}, and \eqref{IoM7}.
        \item $V_0=V_0(r), V_1=V_1(r), V_2=V_2(r)$,
        \[ V_4=0,\quad V_3=\dfrac{\hbar^2+2\hbar r^2 V_1-12r^2 V_2}{6r^4},\quad V_5=-\dfrac{1}{3r^2}+\dfrac{V_1}{3\hbar}. \]
        There are three non-trivial scalar integrals of motion given by \eqref{IoM6}, \eqref{IoM8}, and \eqref{IoM9}.
        \item $V_0=V_0(r), V_1=V_1(r), V_2=V_2(r), V_3=V_3(r)$,
        \[ V_4=0,\quad V_5=-\dfrac{1}{3r^2}+\dfrac{V_1}{3\hbar}. \]
        There is only one non-trivial scalar integral of motion given by \eqref{IoM6}.
        \item $V_0=V_0(r), V_1=V_1(r)$,
        \[ V_4=0,\quad V_5=-\dfrac{1}{3r^2}+\dfrac{V_1}{3\hbar}-\dfrac{2r^2}{\hbar^2}V_3,\quad V_2=\dfrac{\hbar(\hbar+2r^2 V_1)}{12r^2}. \]
        There are three non-trivial scalar integrals of motion given by \eqref{IoM9}, \eqref{IoM19}, and \eqref{IoM20}.
        \item $V_0=V_0(r), V_1=V_1(r), V_2=V_2(r), V_3=V_3(r)$,
        \[ V_4=0,\quad V_5=\dfrac{\hbar V_1-4V_2-2r^2 V_3}{\hbar^2}. \]
        There are two non-trivial scalar integrals of motion given by \eqref{IoM8} and \eqref{IoM9}.
        \item $V_0=V_0(r), V_1=V_1(r), V_2=V_2(r), V_5=V_5(r)$, $V_4=0$, $V_3=0$. There is only one non-trivial scalar integral of motion given by \eqref{IoM7}.
        \item $V_0=V_0(r)$, $V_1=V_1(r)$, $V_2=V_2(r)$,
        \begin{align*}
             &V_4=\dfrac{1}{2},\quad V_5=\alpha_{23},\quad V_3=\dfrac{V_0-3V_2+\alpha_{26}}{r^2}.
        \end{align*}
        There is an infinite family of scalar integrals of motion given by \eqref{IoM21}.
        \item $V_0=V_0(r)$, $V_1=V_1(r)$, $V_2=V_2(r)$, $V_3=V_3(r)$, $V_5=V_5(r)$, and $V_4=\dfrac{1}{2}$. There is an infinite family of scalar integrals of motion given by \eqref{IoM22}.
    \end{enumerate}
    Note that in all the cases listed above $(\vec{\sigma}_1,\vec{\sigma}_2)$ is also scalar integrals of motion since it is a trivial one.
\end{theorem}


In \cite{OM}, Okubo and Marshak pointed out that on the energy shell the potential $V_4$ appearing in the Hamiltonian can be directly derived from the other terms, that is, we can set $V_4=0$. However, off the energy shell, this term must be considered as an independent term. In our analysis, we have considered the most general Hamiltonian including the term $V_4$ independently since the computation was actually not exhausting. However, in our future works, this term $V_4$ might be set to zero to facilitate the analysis.

There are many fruitful research directions we could pursue in the future. We are currently investigating superintegrable systems with spin admitting additional pseudo-scalar integrals of motion, and then we aim to classify all superintegrable systems admitting vector, axial-vector, tensor, and pseudo-tensor integrals of motion. Further exploration of the algebras formed by the integrals derived in this paper is also underway.

\section*{Acknowledgments}
This work was financially supported by TUBITAK (The Scientific and Technological Research Council of T\"{u}rkiye) 1001 program with project number 123F161.

\appendix

\section{Detailed subcase structure for Case 1}
In this appendix, we present a tree diagram that outlines the naming structure of all subcases within Case 1 of the analysis. The diagram includes only the labels of the subcases to provide a clear and concise overview of their organization. This visual aid is intended to help readers navigate the intricate branching of subcases more easily and avoid potential confusion. Note that a similar diagram is not provided for Case 2, as its subcases are straightforward and easy to follow.
\begin{figure}[H]
    \centering
    \includegraphics[width=\linewidth]{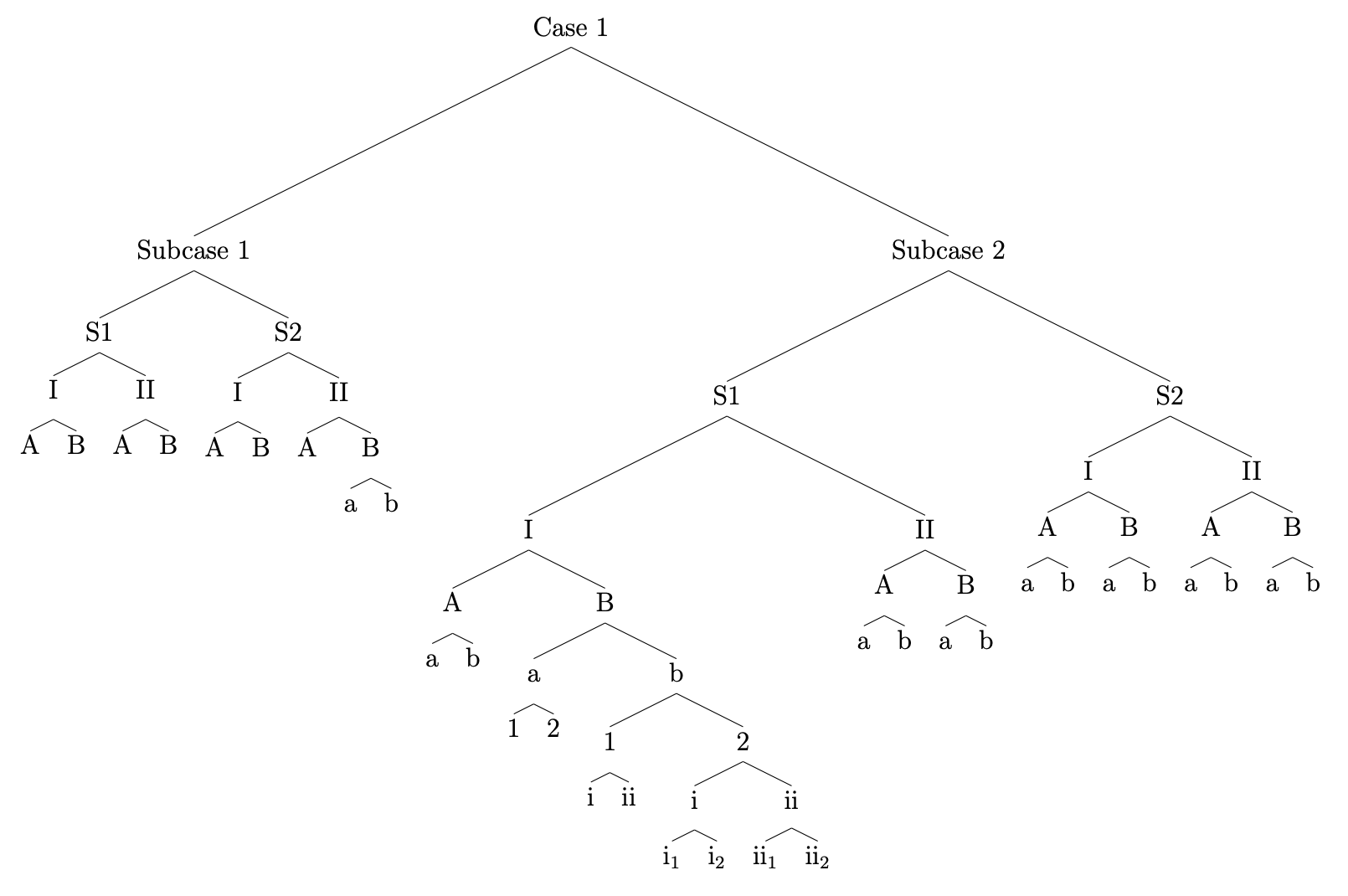}
\end{figure}


\end{document}